\documentclass[journal=jctcce,manuscript=article,layout=onecolumn]{achemso}

\usepackage[version=4]{mhchem}  
\usepackage{dcolumn}            
\usepackage{booktabs}           
\usepackage{bm}                 
\usepackage{upgreek}
\usepackage{hyperref}
\usepackage{graphicx}
\usepackage{epstopdf}
\usepackage{amssymb,mathtools}
\usepackage{amsmath}
\usepackage{tabularx} 
\usepackage{colortbl}
\usepackage{makecell}
\usepackage{multirow}
\usepackage[normalem]{ulem}
\usepackage{enumitem}
\usepackage{color}
\usepackage{xparse}
\usepackage{lmodern}
\usepackage{tikz,xcolor}
\usepackage{mciteplus}
\author{Kinga Warda}
\affiliation{Forschungszentrum Jülich GmbH, Institute of Energy Technologies, Theory and Computation of Energy Materials (IET-3), 52425 Jülich, Germany}
\alsoaffiliation{Jülich Aachen Research Alliance
JARA Energy \& Center for Simulation and Data Science (CSD)
52428 Jülich, Germany}
\alsoaffiliation{Chair of Theory and Computation of Energy Materials Faculty of Georesources and Materials Engineering
RWTH Aachen University, 52062 Aachen, Germany}
\altaffiliation{Contributed Equally}

\author{Eric Macke}
\affiliation{Faculty of Production Engineering, Bremen Center for Computational Materials Science and MAPEX Center for Materials and Processes, Hybrid Materials Interfaces Group, University of Bremen, D-28359 Bremen, Germany}
\alsoaffiliation{U Bremen Excellence Chair, University of Bremen, D-28359 Bremen, Germany}
\altaffiliation{Contributed Equally}

\author{Iurii Timrov}
\affiliation{PSI Center for Scientific Computing, Theory, and Data, Paul Scherrer Institute, 5232 Villigen PSI, Switzerland}

\author{Lucio Colombi Ciacchi}
\affiliation{Faculty of Production Engineering, Bremen Center for Computational Materials Science and MAPEX Center for Materials and Processes, Hybrid Materials Interfaces Group, University of Bremen, D-28359 Bremen, Germany}

\author{Piotr M. Kowalski}
\affiliation{Forschungszentrum Jülich GmbH, Institute of Energy Technologies, Theory and Computation of Energy Materials (IET-3), 52425 Jülich, Germany}
\alsoaffiliation{Jülich Aachen Research Alliance
JARA Energy \& Center for Simulation and Data Science (CSD)
52428 Jülich, Germany}
\email{p.kowalski@fz-juelich.de}

\title{Getting the manifold right: The crucial role of orbital resolution in DFT\texorpdfstring{$+\boldsymbol{U}$}{+U} for mixed \textit{d–f} electron compounds}

\abbreviations{DFT+U,DFT,GGA,PBEsol,LDA,OAO,NAO,WF,TM,REE,PBEsol,OR-DFT+U}

\begin{document}

\begin{tocentry}

    \centering
    \includegraphics[width=8.3cm]{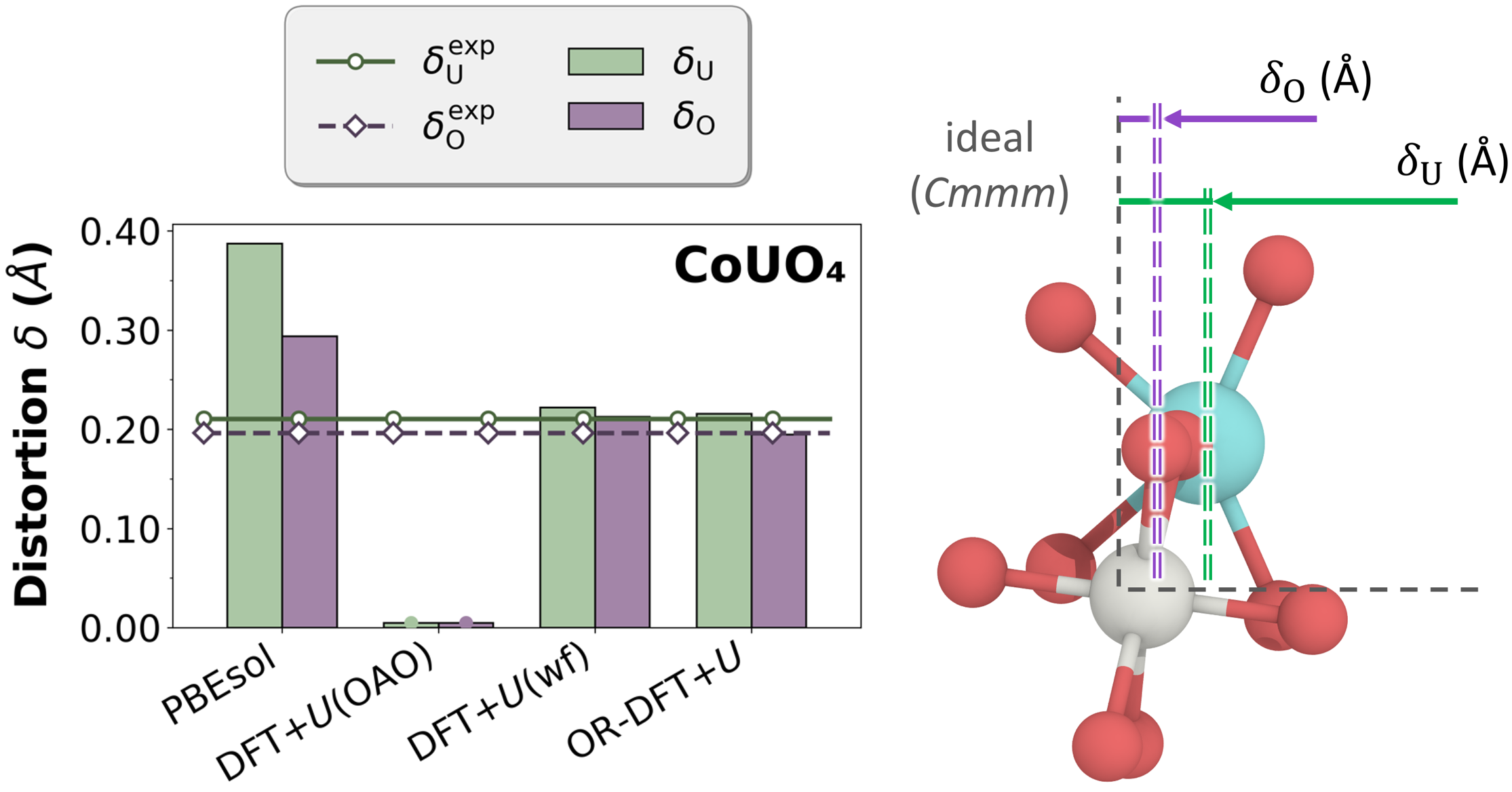}

\end{tocentry}


\newpage
\begin{abstract}
Accurately modeling compounds with partially filled $d$ and $f$ shells remains a hard challenge for density-functional theory, due to large self-interaction errors stemming from local or semi-local exchange-correlation functionals.
Hubbard $U$ corrections can mitigate such errors, but are often detrimental to the description of hybridized states, leading to spurious force contributions and wrong lattice structures.
Here, we show that careful disentanglement of localized and delocalized states leads to accurate predictions of electronic states and structural distortions in ternary monouranates (\ce{AUO4}, where A represents Mn, Co, or Ni), for which standard $U$ corrections generally fail.
Crucial to achieving such accuracy is a minimization of the mismatch between the spatial extension of the projector functions and the true coordination geometry.
This requires Wannier-like alternatives to atomic-orbital projector functions, or corrections of Hubbard manifolds exclusively comprised of the most localized A-$3d$, U-$5f$ and O-$2p$ orbitals.
These findings open up the computational prediction of fundamental properties of actinide solids of critical technological importance.  
\end{abstract}
\newpage
\section{Introduction}
    \label{sec:introduction}
Computational investigations play a crucial role in the research workflow on energy materials~\cite{jainComputational2016}.
This is especially the case for nuclear materials, where direct measurements are often limited to surrogate systems in which hazardous actinide elements are replaced by harmless lanthanides~\cite{chroneosNuclearWasteformMaterials2013, bosbachResearchSafeManagement2020}.
However, the key elements that determine the performance of electrocatalysts, batteries or nuclear fuel and waste, featuring partially filled $d$ and $f$ shells – including transition metals and actinides – pose a challenge for density-functional theory (DFT), the main tool for prediction of the structure and electronic properties of functional materials.
This is because the (typically localized) $d$ and $f$ electrons are particularly prone to spurious self-interaction errors (SIEs) that arise from the inexact cancellation of the Hartree term by approximate exchange-correlation (xc) functionals~\cite{perdewSelfinteractionCorrection1981a,mori-sanchez_many-electron_2006,kummelOrbitaldependentDensityFunctionals2008}.
As a result, several features of exact DFT are not correctly reproduced by LDA and GGA-type xc functionals: the piecewise linearity (PWL) of the total energy with respect to fractional addition or removal of charge,~\cite{perdewSelfinteractionCorrection1981a,bajaj_communication_2017,burgess_dft_2023} the existence of a derivative discontinuity,~\cite{perdewRestoringDensityGradientExpansion2008a,zhao_global_2016} and the correct asymptotic decay of the Kohn-Sham (KS) potential~\cite{kronikPiecewiseLinearityFreedom2020}.
These shortcomings ultimately compromise the prediction of several fundamental properties, including lattice symmetries, bond strengths and band gaps.
For instance, GGA functionals predict \ce{UO2} ---the main form of nuclear fuel--- to be metallic, whereas it is in fact a Mott insulator with a band gap of $\sim 2\, \rm eV$~\cite{beridzeBenchmarkingDFTMethod2014}.

One way to mitigate SIEs is the DFT$+U$ framework~\cite{anisimov_band_1991,solovyev_corrected_1994, liechtenstein_density-functional_1995, dudarevElectronenergylossSpectraStructural1998}.
In this approach, a corrective on-site term scaled by a Hubbard parameter $U$ is applied to a predefined electronic subspace dubbed Hubbard manifold in order to remove (locally) the spurious quadratic deviations from PWL,~\cite{cococcioniLinearResponseApproach2005a} while also reintroducing a derivative discontinuity.
Note that due to its history as a remedy to the unsatisfactory performance of (semi-)local functionals in predicting the electronic structure of strongly correlated materials, DFT$+U$ is often understood as a method that corrects for strong electronic correlations not well captured by DFT; however, this view has been increasingly challenged~\cite{cococcioniLinearResponseApproach2005a,kulikDensityFunctionalTheory2006,mackeOrbitalResolvedDFTMolecules2024}.
In the commonly used approximation, the Hubbard manifold encompasses the entirety of valence $d$ and $f$ shells, and all magnetic quantum orbitals of these shells are corrected using a \textit{shell-averaged} scalar Hubbard $U$ parameter.
This approximation implicitly relies on the assumption that all $d$ and $f$ states are localized and therefore display only weak interactions with the surrounding electron bath.
To apply the correction, the orbital occupations within the Hubbard manifold are determined by projecting the Kohn–Sham wavefunctions onto a suitable mathematical representation, henceforth referred to as Hubbard projectors.
A straightforward and widely used choice for this representation consists in atomic-like orbitals parameterized based on solutions of radial Schrödinger equations for isolated atoms.
Shell-averaged DFT$+U$ with such atomic orbital projectors has been successfully applied to study the electronic and ionic structures of a variety of solids bearing lanthanides and actinides~\cite{chenDFTStudyUranium2022,anderssonDensityFunctionalTheory2013,kvashninaTrendsValenceBand2018,Romero2014}.

Geological disposal of nuclear waste is a challenge faced by many countries that utilize nuclear technology~\cite{ewingGeologicalDisposalNuclear2016}.
A particular concern is the formation of secondary phases between disposed \ce{UO2}-based nuclear fuel, ﬁssion products, and near-field elements. 
A notable example of these secondary phases is provided by ternary monouranates \ce{AUO4}, with \ce{A} representing a bi- or trivalent metal~\cite{murphySolidState2019}.
Such materials are often studied using DFT+$U$ schemes.
However, for \ce{AUO4} compounds with \ce{A=Ni,Mg,Co,Mn}, the shell-averaged approach was shown to incorrectly stabilize a higher-symmetry $Cmmm$ structure instead of the experimentally observed $Ibmm$ phase~\cite{murphyTiltingDistortionRutileRelated2021}.
The failure of shell-averaged DFT$+U$ was attributed to a significant overestimation of $d$ and $f$ orbital occupancies due to the use of atomic orbital Hubbard projectors, which led to large artificial Hubbard energy contributions~\cite{murphyTiltingDistortionRutileRelated2021}.
To demonstrate this, the authors set the Hubbard energy term $E_U$ to zero, and with this recovered the experimentally observed $Ibmm$ structures.
Only for \ce{A=Cd} did the $Cmmm$ structure remain stable, again in line with experimental results.
Improvements similar to the $E_U=0$ procedure were obtained by replacing the atomic-like Hubbard projectors with Wannier-type ones (hereafter denoted as the DFT+$U$(WF) method).
Wannier-type Hubbard projectors yield more realistic (i.e., closer to integer values) $d$ and $f$ occupations, so that the spurious Hubbard energy contributions vanish naturally.
The DFT+$U$(WF) method has also proven valuable in other systems containing actinides and transition-metal (TM) atoms, offering substantial improvements in the predicted electronic structure, thermodynamic properties, and X-ray spectra~\cite{kvashninaTrendsValenceBand2018,tingRefinedDFTMethod2023,heLowspinStateFe2023,vitovaDehydrationUranylPeroxide2018,kowalskiElectrodeElectrolyteMaterials2021}. 

Despite these advantages, DFT$+U$(WF) comes with practical limitations: the construction of Wannier functions is sensitive to initialization choices (energy windows, number of bands, disentanglement procedure), and---depending on the type of Wannier function used---the direct evaluation of forces and stresses is not yet implemented in the codes.
An alternative approach that also seeks to avoid the overcorrections of shell-averaged DFT$+U$ is orbital-resolved DFT$+U$ (OR-DFT$+U$)~\cite{mackeOrbitalResolvedDFTMolecules2024}.
This method is grounded in the insight that the degree of electron localization (and thus the severity of SIEs) can vary significantly at the \textit{intra-shell} level, i.e., between different $nlm$ orbitals of the same $nl$ subshell~\cite{zhouObtainingCorrectOrbital2009}.
OR-DFT$+U$ provides an \textit{ad-hoc} solution that enables the definition of pinpoint Hubbard manifolds spanned by simple atomic orbital projectors for compounds where shell-averaged DFT+$U$ fails (Refs.~\citenum{solovyev2All31996,pickettReformulationLDAMethod1998,oregan_subspace_2011,mariano_biased_2020,mackeOrbitalResolvedDFTMolecules2024} report examples of such failures).

In this study, we apply various orbital-resolved Hubbard manifolds to the ternary TM monouranates \ce{\beta-NiUO4}, \ce{CoUO4} and \ce{MnUO4}, aiming to reproduce the experimentally observed structural distortions without imposing additional constraints on the Hubbard energy or introducing other empirical assumptions or modifications.
All Hubbard $U$ parameters are computed from first principles using the linear-response constrained DFT approach (LR-cDFT)~\cite{pickettReformulationLDAMethod1998,cococcioniLinearResponseApproach2005a}.
By testing different orbital-resolved Hubbard manifolds (determined through careful analyses of orbital occupations), we aim at identifying the root cause of the failure of shell-averaged DFT$+U$ in these systems.
Finally, we assess how the definition of the Hubbard manifold and the choice of projector functions influence the prediction of structural observables and clarify under which conditions OR-DFT$+U$ can serve as a robust and practical alternative to the DFT$+U$(WF) framework for modeling $d$ and $f$ electron systems with strong covalent bonding.

\section{Materials and methods}
\subsection{Computational details}
    \label{sec:details}
DFT calculations were performed using the \texttt{pw.x} code contained in the \textsc{Quantum ESPRESSO} package~\cite{giannozziQUANTUMESPRESSOModular2009b, giannozziAdvancedCapabilitiesMaterials2017b, Giannozzi:2020}.
The calculation setup was chosen to match the computational setup used by \citeauthor{murphyTiltingDistortionRutileRelated2021}
This included employing the PBEsol exchange-correlation functional~\cite{perdewRestoringDensityGradientExpansion2008a} and ultrasoft pseudopotentials (self-generated using the Vanderbilt code~\cite{vanderbiltSoftSelfconsistentPseudopotentials1990}) with a plane-wave energy cutoff of 50\,Ry and a charge density cutoff of 200\,Ry. 
Structural relaxations were carried out until the total energy and atomic forces converged below 10$^{-5}$\,Ry and 10$^{-4}$\,Ry/Bohr, respectively.
All calculations were spin-polarized within the collinear approximation (assuming ferromagnetic order of the TM sites) and used a tight convergence threshold for electronic total-energy self-consistency of $10^{-8}\,$Ry.
Projected density of states (PDOS) calculations were performed with a Gaussian broadening of 0.0035\,Ry.
Moreover, the option \texttt{diag\_basis\,=\,.true.} in the \texttt{projwfc.x} code was enabled to rotate atomic orbitals into the eigenbasis of the occupation matrix, thus aligning orbital projections with the local symmetry and facilitating a clearer identification of orbital character~\cite{mahajanImportance2021}.
An additional analysis of the electronic structure of \ce{NiUO4} employed the linear combination of fragment orbitals method~\cite{mullerFragment2024} as implemented in the \textsc{LOBSTER} tool~\cite{maintzLOBSTER2016}, for which a single calculation had to be re-done using PAW-type pseudopotentials (taken from the PSlibrary).

For all DFT$+U$ calculations, we applied the formulation of \citeauthor{dudarevElectronenergylossSpectraStructural1998} that seeks to mitigate SIEs by enforcing (locally) PWL of the total energy with respect to fractional orbital occupations~\cite{cococcioniLinearResponseApproach2005a,kulikDensityFunctionalTheory2006}.
Note that in this work, ``DFT+$U$'' refers exclusively to PBEsol+$U$.
We used both shell-averaged and orbital-resolved Hubbard manifolds~\cite{solovyev2All31996,pickettReformulationLDAMethod1998,mackeOrbitalResolvedDFTMolecules2024}.
The OR-DFT$+U$ energy functional is given by
\begin{equation}
    E_{\mathrm{DFT}+U} = E_{\mathrm{DFT}} + E_U \,,
    \label{eq:edftu}
\end{equation}
where
\begin{equation}
    E_U = \sum_{I,\sigma} \sum_{i=1}^{2l+1} \frac{U_i^I}{2} \lambda_{i}^{I\sigma}(1-\lambda_{i}^{I\sigma}) \,,
    \label{eq:Ehub}
\end{equation}
and the corresponding Hubbard potential is
\begin{equation}
    V_{U}^{\sigma}=  \sum_I \sum_{i=1}^{2l+1} U^I_i \left(\frac{1}{2} -  \lambda^{I\sigma}_{i} \right)|\phi^I_i \rangle \langle \phi^I_{i}| .
    \label{eq:hubbard-potential-u}
\end{equation}
Here, $U_i^I$ is the Hubbard parameter corresponding to the (diagonal) orbital $\{\phi_i^I\}$ of atom $I$ (the principal and angular quantum numbers \textit{nl} are omitted for clarity), $\sigma$ is the spin index, and $\lambda^{I\sigma}_i$ are the eigenvalues of the occupation matrix, computed by solving the eigenvalue problem $\sum_{m'} n^{I\sigma}_{m m'} \, \nu^{I\sigma}_{m', i} = \lambda^{I\sigma}_i \, \nu^{I\sigma}_{m, i}$, where $\nu^{I\sigma}$ are the corresponding eigenvectors and $n^{I\sigma}_{m m'}$ are the elements of the occupation matrix for the magnetic quantum numbers $m$ and $m'$. 
In the shell-averaged case, both expressions simplify as $U_i^I = U^I \textrm{ for all } i $. 
Orthogonalized atomic orbitals (OAO)$\{\varphi^I_m\}$ were employed as Hubbard projector functions.
In Eq.~\ref{eq:hubbard-potential-u}, these orbitals are written in their diagonal basis $\{\phi_i^I\}$, which can be back-transformed into the non-diagonal setting by substituting $|\phi_i^{I\sigma}\rangle=\sum_m \nu^{I\sigma}_{mi} \, |\varphi^I_m\rangle$.
We highlight that the Hubbard projector functions are independent of spin; however, for each spin channel, they are rotated separately using the corresponding eigenvectors of the spin-resolved occupation matrix.

OAO are closely related to the nonorthogonalized projectors used by \citeauthor{murphyTiltingDistortionRutileRelated2021}, but prevent the double-counting of Hubbard corrections in regions where projector orbitals overlap~\cite{lowdinNonOrthogonalityProblemConnected1950,timrovPulayForcesDensityfunctional2020}.
With these projector functions, the occupation matrix $\textbf{n}$ is computed as 
\begin{equation}
    {n}^{I\sigma}_{mm'} = \sum_{\mathbf{k},v}
    f_{\mathbf{k},v}^\sigma
    \langle \psi_{\mathbf{k},v}^{\sigma}|\varphi^I_{m'}\rangle \langle \varphi^I_{m}|\psi_{\mathbf{k},v}^{\sigma} \rangle \, ,
    \label{eq:occupations}
\end{equation}
where $m$ and $m'$ are magnetic quantum numbers, and $\psi_{\mathbf{k},v}^\sigma$ are the KS wavefunctions at point $\mathbf{k}$ with the band index $v$ and spin $\sigma$ (additional operators arise when using ultrasoft and PAW basis sets~\cite{timrovPulayForcesDensityfunctional2020}).
All on-site $U$ parameters were evaluated from first principles using the LR-cDFT approach introduced in Ref.~\citenum{cococcioniLinearResponseApproach2005a} and extended to the orbital-resolved formalism in Ref.~\citenum{mackeOrbitalResolvedDFTMolecules2024}.
It is important to note that shell-averaged Hubbard $U$ values are not equivalent to the arithmetic mean of the orbital-resolved $U$ parameters associated with a given shell~\cite{mackeOrbitalResolvedDFTMolecules2024}.
Instead, the shell-averaging is performed at level of the response to the perturbation, which frequently results in shell-averaged $U$ values exceeding both the individual orbital-resolved values and their arithmetic mean~\cite{pickettReformulationLDAMethod1998}. 
In the LR-cDFT calculations, perturbations of magnitude $\alpha = \pm0.05\rm \,eV$ were applied to the respective manifold of a single atom in a $2\times2\times2$ supercell that was created to avoid unphysical interactions of the perturbed state with its periodic images.
It is important to note that the \textsc{Quantum ESPRESSO} implementation of LR-cDFT ensures that only the DFT part of the linear response is assessed, as the potential derived from the Hubbard functional is kept fixed during perturbative calculations~\cite{cococcioniEnergetics2019, Timrov:2018}.
The Brillouin zone was sampled using $\Gamma$-centered $4\times4\times4$ and $2\times2\times2$ Monkhorst–Pack grids for the unit cell and the supercell calculations, respectively.
As the responses to perturbations depend on the electronic structure (which in turn depends on the ionic structure), calculated Hubbard parameters can change when transitioning from a DFT ground state to a DFT$+U$ one.
While there exist procedures that self-consistently derive the Hubbard parameters and optimize the ionic structure~\cite{cococcioniEnergetics2019,timrovSelfconsistentHubbardParameters2021}, we did not adopt such an approach here.
Instead, all LR-cDFT calculations were carried out in a ``one-shot'' manner on top of DFT$+U$ ground states obtained using an empirical guess for the $U$ values: $U_{\mathrm{Ni\text{-}3}d} = U_{\mathrm{Co\text{-}3}d}=4.0\,$eV, $U_{\mathrm{Mn\text{-}3}d}=2.0\,$eV, $U_{\mathrm{U\text{-}5}f}=2.0\,$eV, and $U_{\mathrm{O\text{-}2}p}=1.0\,$eV.
This procedure was previously shown to yield Hubbard parameters typically within $\delta U = U_{\mathrm{SC}} - U\leq 0.1\,$eV (where $U_{\mathrm{SC}}$ is the self-consistent value), provided the input structure does not deviate strongly from the final relaxed one~\cite{bastoneroFirstprinciplesHubbardParameters2025}.
We have confirmed that this rapid convergence also holds for the \ce{AUO4} compounds discussed in the following (Table S3).

\subsection{Characterization of the structural distortions}
\label{subsec:distortions-formulae}
\begin{figure*}[hb!]
    \centering
    \includegraphics[width=1.0\textwidth]{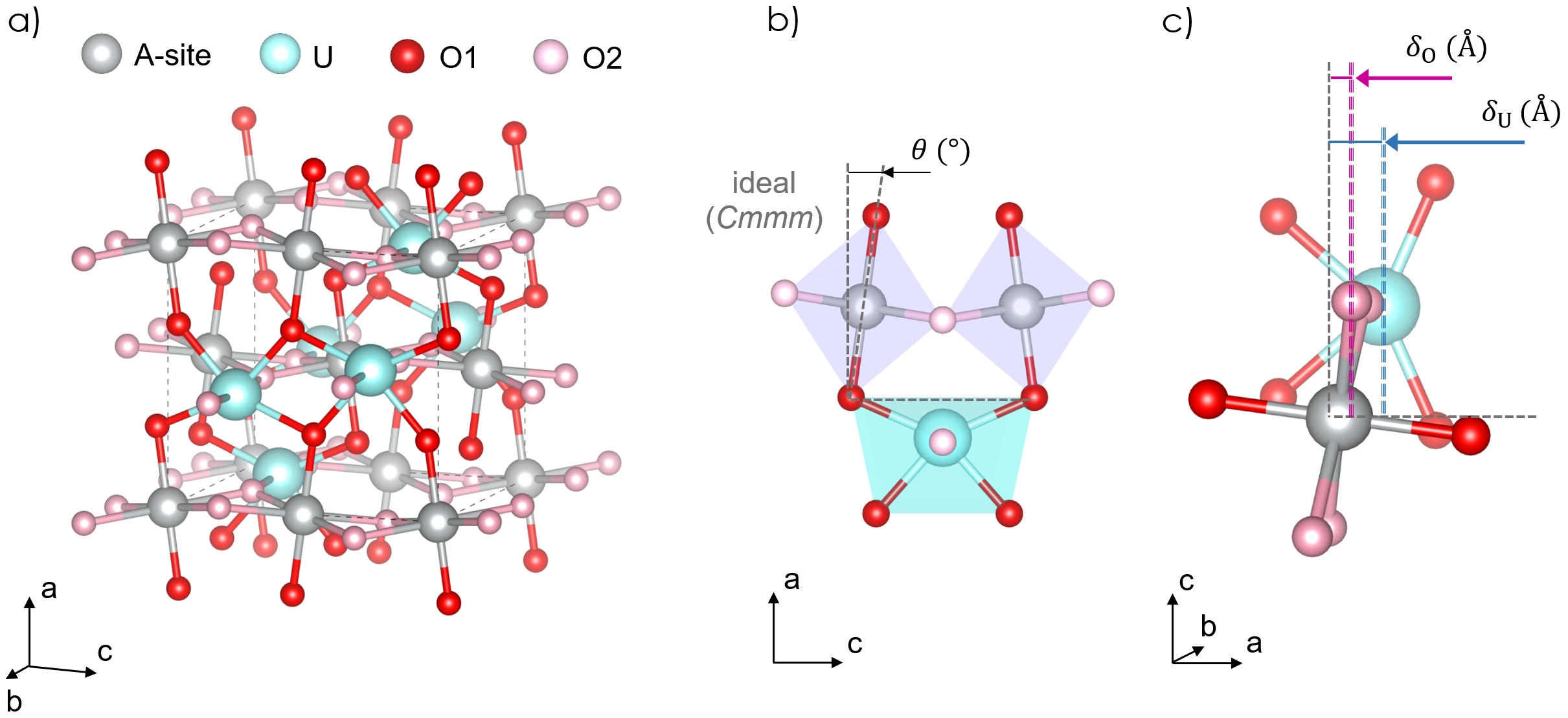}
    \caption{(a) Crystal structure of \ce{AUO4} in the orthorhombic \textit{Ibmm} space group. Oxygen atoms O1 and O2 are shown in red and pink, respectively. (b) Schematic representation of the \ce{AO6} tilt angle $\theta$ around the b-axis, quantifying deviations from the idealized \textit{Cmmm} symmetry. The \ce{UO6} octahedra is shown in cyan and the \ce{AO6} octahedra in purple. (c) The uranium off-centering $\delta_\mathrm{U}$ and axial oxygen off-centering $\delta_\mathrm{O}$, measured relative to the ideal $Cmmm$ atomic positions. The high-symmetry $Cmmm$ reference structure is indicated by dashed gray lines.} 
    \label{fig:structure}
\end{figure*}

The crystal structure of $Ibmm$ ternary monouranates is similar to rutile (\ce{TiO2}, space group $P\,{4_2}/{m}\,n\,m$)~\cite{baurCoReO4NewRutiletype1992}, featuring parallel one-dimensional chains of edge-sharing \ce{AO6} and \ce{UO6} octahedra that run along the $[001]$ direction (Figure~\ref{fig:structure}a). 
Each \ce{U} atom is coordinated by two O1 atoms in a \textit{trans} configuration and four equatorial O2 atoms, whereas the reverse arrangement occurs in the \ce{AO6} octahedra. 
A distinctive feature of the $Ibmm$ structure are strongly hybridized uranyl \ce{O$1$=U=O$1$} moieties, which display significantly shorter bond lengths than the \ce{U-O$2$} bonds~\cite{baurCoReO4NewRutiletype1992}.
These moieties are associated with structural distortions~\cite{murphyHighPressureSynthesisStructural2018}---namely octahedral tilting ($\theta$, Figure~\ref{fig:structure}b) and off-centering distortions of uranium and oxygen atoms ($\delta$, Figure~\ref{fig:structure}c)---with respect to the higher-symmetry $Cmmm$ structure of \ce{CdUO4}.  
The tilting angle $\theta$ measures the deviation of the O2 atoms from collinearity with the b-axis and is given by  
\begin{equation}
    \theta = 90^\circ - \tan^{-1} \left[ \frac{(2x - 1) a}{2zc} \right],
    \label{eq:placeholder_label}
\end{equation}
where $x$ and $z$ are the fractional coordinates of the O2 atoms, and $a$ and $c$ are lattice parameters.  
The uranium off-centering $\delta_\mathrm{U}$ and the oxygen off-centering $\delta_\mathrm{O}$ are defined as the displacements of the U and O2 atoms from the center of the adjacent \ce{AO6} octahedron, measured along the [100] direction (Figure~\ref{fig:structure}c). 
Specifically, $\delta_\mathrm{U}$ and $\delta_\mathrm{O}$ correspond to the differences between the atomic coordinates of U and O2, and the coordinate of the octahedral center, respectively, all projected onto the [100] direction, capturing the structural distortion along the primary axis of asymmetry.
Displacements along other directions are neglected.

\section{Results and discussion}
\subsection{Response of the structure to variations in \texorpdfstring{$U$}{U}}
\begin{figure}[ht!]
    \centering
    \includegraphics[width=0.5\textwidth]{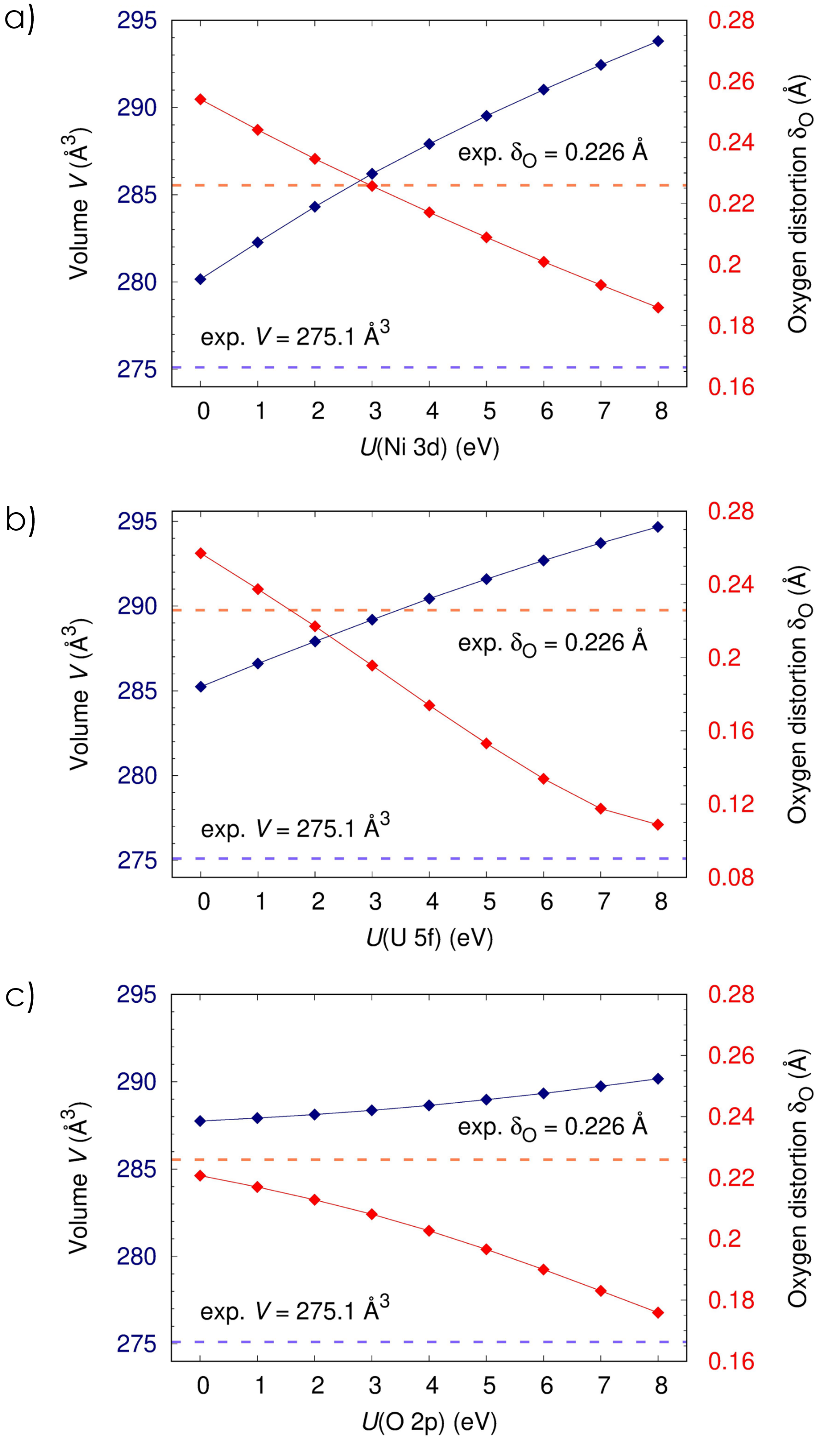}
    \caption{Impact of the Hubbard parameters (a)~$U_{\mathrm{Ni\text{-}3}d}$ (at fixed $U_{\mathrm{U\text{-}5}f} = 2$\,eV and $U_{\mathrm{O\text{-}2}p} = 1$\,eV), (b)~$U_{\mathrm{U\text{-}5}f}$ (at fixed $U_{\mathrm{Ni\text{-}3}d} = 4$\,eV and $U_{\mathrm{O\text{-}2}p} = 1$\,eV), and (c)~$U_{\mathrm{O\text{-}2}p}$ (at fixed $U_{\mathrm{Ni\text{-}3}d} = 4$\,eV and $U_{\mathrm{U\text{-}5}f} = 2$\,eV fixed) on the cell volume and the magnitude of the oxygen distortion parameter in \ce{\beta-NiUO4}. Solid lines serve as a guide for the eye.}
    \label{fig:u-on-niuo4}
\end{figure}

To test how \ce{AUO4} structures responds to shell-averaged Hubbard $U$ corrections, we performed geometry optimizations on \ce{\beta-NiUO4} in which the values of $U_{\mathrm{Ni\text{-}3}d}$,  $U_{\mathrm{U\text{-}5}f}$, and $U_{\mathrm{O\text{-}2}p}$ were varied systematically.
This compound was selected for the analysis because it displays the most distorted structure.
The results, presented in Figure~\ref{fig:u-on-niuo4}, show a strong and monotonic dependence of both the unit cell volume $V$ and the internal distortion parameter $\delta_\mathrm{O}$ on the applied $U$ values.
Specifically, increasing $U_{\mathrm{Ni\text{-}3}d}$ from 0 to 8\,eV  results in a 4.8\% expansion of the volume and a 27\% reduction in $\delta_\mathrm{O}$ (Figure~\ref{fig:u-on-niuo4}a).
Similarly, increasing $U_{\mathrm{U\text{-}5}f}$ from 0 to 8\,eV leads to a 3.3\% volume expansion and a pronounced reduction in $\delta_\mathrm{O}$ by 58\% (Figure~\ref{fig:u-on-niuo4}b), which reflects the crucial role of the U-$5f$ orbitals for the strongly covalent hybridized axial \ce{O$1$=U=O$1$} bonds. 
In contrast, applying $U$ corrections to the ${\mathrm{O\text{-}2}p}$ shell produces a subtler structural response, as increasing $U_{\mathrm{O\text{-}2}p}$ from 0 to 8\,eV entails only a minimal increase in unit cell volume (0.8\%) and a reduction in $\delta_\mathrm{O}$ by 20\% (Figure~\ref{fig:u-on-niuo4}c).
The experimental value of $\delta_\mathrm{O}$ is best reproduced at $U_{\mathrm{Ni\text{-}3}d} \approx 3$\,eV, $U_{\mathrm{U\text{-}5}f} \approx 1.5$\,eV and $U_{\mathrm{O\text{-}2}p} \approx 0$\,eV, whereas the volume is overestimated for all $U$ values, including 0\,eV. 
Qualitatively similar trends are observed for \ce{MnUO4} and \ce{CoUO4}, suggesting that the $Ibmm$ structure of \ce{AUO4} compounds manifests a high sensitivity to shell-averaged Hubbard $U$ corrections; at least when the latter are applied through (ortho-)atomic projector functions.
For ranges of Hubbard $U$ values that predict correct electronic band gaps (e.g., $U>6\,$eV for Ni-$3d$)~\cite{uhrinMachineLearningHubbard2025} the magnitude of the distortions is severely underestimated, especially since the effects of $U_{\mathrm{Ni\text{-}3}d}$,  $U_{\mathrm{U\text{-}5}f}$, and $U_{\mathrm{O\text{-}2}p}$ can be expected to stack (albeit not necessarily in a linear way).

\citeauthor{murphyTiltingDistortionRutileRelated2021} explained this inconsistent behavior of the $U$ correction with ``unrealistic'' fractional occupation numbers of unoccupied $d$ and $f$ orbitals resulting from the use of atomic orbital projectors~\cite{murphyTiltingDistortionRutileRelated2021}.
Due to the functional form of Eq.~\ref{eq:Ehub}, fractional occupation numbers induce punitive Hubbard energy contributions that grow quadratically with increasing distance from $\lambda=0$ or $\lambda=1$ (i.e., $E_U$ is maximized for half-occupied orbitals).
Therefore, the minima of the potential energy surface shift towards configurations in which $E_U$ is reduced.
For (ortho)-atomic projector functions, this entails the observed volume increase---since larger bond lengths lead to decreased occupations of formally empty orbitals---and causes a linearization of the polyhedral bond angles that reduces the magnitudes of $\delta_U$ and $\delta_O$.
In Ref.~\citenum{murphyTiltingDistortionRutileRelated2021} these spurious side-effects of $U$ corrections were mitigated by using WFs as Hubbard projectors, which yielded occupation numbers much closer to zero or one, or by setting $E_U=0$ when using atomic Hubbard projectors.
The fact that the problematic contributions to $E_U$ appear to stem from only a few (formally unoccupied) states raises the question of whether there exists an orbital-resolved Hubbard manifold based on (ortho\mbox{-)}atomic orbital projectors that reduces the sensitivity of the structural parameters to the $U$ values and allows for accurate predictions of the structural distortions without resorting to artificial modifications of the energy functional.

\subsection{Determination of orbital-resolved Hubbard manifolds}
    \label{subsec:manifolds}
To identify such orbital-resolved Hubbard manifolds, we analyze the individual orbitals of the \ce{A}-$3d$, \ce{U}-$5f$ and \ce{O}-$2p$ shells in terms of overlap and occupation patterns.
We emphasize that the consideration of \ce{O}-$2p$ states for the Hubbard manifold is natural because of their potential frontier-state character and high degree of localization~\cite{orhanFirstprinciplesHubbardHunds2020}.
\subsubsection{Group-theoretical analysis of orbital splitting}
Before turning to practical calculations, we review the ligand-field-induced splitting of the \ce{A}-$3d$ and \ce{U}-$5f$ orbitals from a group theory perspective.
In the $Ibmm$ crystal structure, the \ce{A} and \ce{U} cations occupy the 4a and 4e Wyckoff positions, corresponding to $C_{2h}$ and $C_{2v}$ site symmetries, respectively.
The \ce{A}-$3d$ orbitals transform as two one-dimensional (non-degenerate) irreducible representations: a $3A_g$ representation with contributions from $d_{x^2-y^2}$, $d_{z^2}$, and $d_{xz}$, and a $2B_g$ manifold associated with the $d_{xy}$ and $d_{yz}$ atomic orbitals. 
The lobes of $d_{xy}$, $d_{xz}$ and $d_{z^2}$ lie between the \ce{A-O} bond axes and should interact only very weakly with the O-$2p$ orbitals via $\pi$ overlap (Figure~S7).
In contrast, the lobes of $d_{x^2-y^2}$ and $d_{yz}$ run along the equatorial and axial \ce{A-O} bonds, respectively, giving rise to strong $\sigma$ overlap.
Therefore, despite the significant distortions of the \ce{AO6} octahedra, the crystal-field splitting pattern of the \ce{A}-$3d$ shell can be expected to resemble that of a perfect octahedron with $O_h$ site symmetry.
We will thus refer to the three lowest-lying orbitals as $\widetilde{t_{2g}}$, and we designate the two higher-energy orbitals as $\widetilde{e_g}$, where the tilde denotes the approximate nature of this assignment, since all of the states are non-degenerate in energy.

Moving on, the $5f$ orbitals of \ce{U} split into seven non-degenerate levels transforming as $2A_1$, $1A_2$, $2B_1$, and $2B_2$. 
The $A_2$ representation can be expected to show very little overlap, as the lobes of the corresponding $f_{xyz}$ orbital point away from all bond axes (Figure~S7).
All other representations are linear combinations of two atomic orbitals and exhibit varying degrees of overlap with adjacent \ce{O}$-sp^2$ hybrid orbitals; with the highest energy contributions likely being due to $f_{y(3x^2-y^2)}$ and $f_{xz^2}$, whose lobes follow the axial \ce{\text{O1}=U=\text{O1}} and the equatorial \ce{\text{O2}-U-\text{O2}} bond axes, respectively.

\subsubsection{Analysis of the electronic occupations}
\label{subsec:trial-calc}
\begin{table}[ht]
    \fontsize{8pt}{8pt}\selectfont
    \centering
    \setlength{\tabcolsep}{4pt} 
    \renewcommand{\arraystretch}{1.4} 
    \caption{Eigenvalues $\lambda_i^I$ corresponding to respective eigenstates $\nu_i^I$ in \ce{\beta-NiUO4}, obtained from a DFT+$U$ calculation with trial Hubbard parameters $U_{\mathrm{Ni\text{-}3}d}$ = 4.0\,eV, $U_{\mathrm{U\text{-}5}f}$ = 2.0\,eV and $U_{\mathrm{O\text{-}2}p}$ = 1.0\,eV.}
    \label{tab:eigenvalues}
    \begin{tabular}{c c c c c c c c c}
        \arrayrulecolor{black}\hline
          Atom & Spin &  \multicolumn{7}{c}{Eigenvalue} \\   
          &  & $\lambda_1$ & $\lambda_2$ & $\lambda_3$ & $\lambda_4$ & $\lambda_5$ & $\lambda_6$ & $\lambda_7$\\ \hline
          U & $\uparrow$ & 0.063 & 0.107 & 0.175 & 0.196 & 0.247 & 0.285 & 
          0.417\\
          & $\downarrow$ & 0.062 & 0.108 & 0.150 & 0.165 & 0.202 & 0.241 & 0.378\\
        \arrayrulecolor{gray}\hline
          Ni & $\uparrow$ & 0.972 & 0.989 & 0.992 & 0.994 & 0.996 & - & -\\
          & $\downarrow$ & 0.164 & 0.221 & 0.983 & 0.987 & 0.991 & - & -\\
        \hline
          O$_{1}$ & $\uparrow$ & 0.748 & 0.765 & 0.812 & - & - & - & -\\
          & $\downarrow$ & 0.736 & 0.766 & 0.810 & - & - & - & -\\
        \hline
          O$_{2}$ & $\uparrow$ & 0.744 & 0.799 & 0.802 & - & - & - & -\\
          & $\downarrow$ & 0.733 & 0.759 & 0.799 & - & - & - & -\\
        \arrayrulecolor{black}\hline
    \end{tabular}
\end{table}
Next, we assess the extent of orbital hybridization by carrying out DFT+$U$ calculations with shell-averaged trial $U$ values and analyze the resulting electronic structures and orbital occupations.
For all trial calculations, we employed $U_{\mathrm{U\text{-}5}f} = 2.0$\,eV and $U_{\mathrm{O\text{-}2}p} = 1.0$\,eV.
The Hubbard parameters used for the \ce{A}-site $3d$ shells were system-specific: $U_{\mathrm{Ni\text{-}3}d} = 4.0$\,eV, $U_{\mathrm{Mn\text{-}3}d} = 2.0$\,eV, and $U_{\mathrm{Co\text{-}3}d} = 4.0$\,eV.
These comparatively low trial parameters were chosen to avoid analyzing an already overcorrected electronic structure, as spurious effects due to shell-averaged Hubbard $U$ corrections often set in at $U>4\,$eV~\cite{mackeOrbitalResolvedDFTMolecules2024,liuAnomalousReversalStability2025}.
The parameter $U_{\mathrm{Mn\text{-}3}d}$ was assigned a smaller value than $U_{\mathrm{Ni\text{-}3}d}$ and $U_{\mathrm{Co\text{-}3}d}$ because previous studies indicate that the \ce{Mn}-$3d$ shell is often less affected by SIEs than \ce{Ni}-$3d$ and \ce{Co}-$3d$~\cite{uhrinMachineLearningHubbard2025}.

\begin{figure*}[ht!]
\centering
     \includegraphics[width=0.9\textwidth]{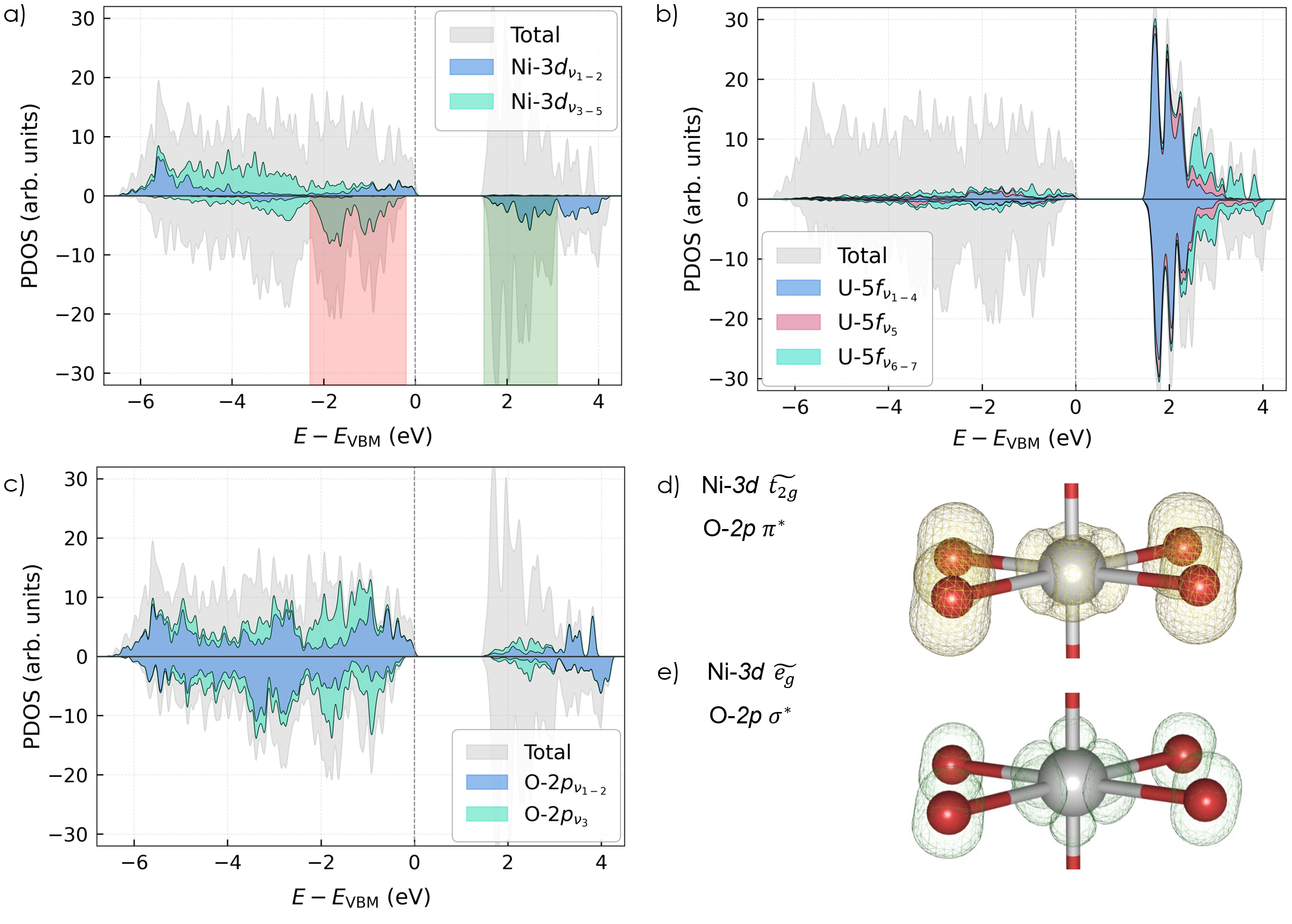}
      \caption{Stacked projected density of states (PDOS) for (a)~Ni-$3d$, (b)~U-$5f$, and (c)~O-$2p$ orbitals in \ce{\beta-NiUO4}, obtained from trial DFT+$U$ calculations. Panels (d) and (e) show the integrated local density of states for the Ni-$3d$ energy intervals indicated by red and green shading in panel (a) at isovalues 0.0184 $e^-\cdot $\AA$^{-3}$ and 0.0079 $e^-\cdot $\AA$^{-3}$, respectively.}
       \label{fig:pdos-isosurfaces}
\end{figure*}

The eigenvalues of the occupation matrix, presented in Table~\ref{tab:eigenvalues}, clearly reflect the crystal-field splitting between the $\widetilde{e_g}$ and $\widetilde{t_{2g}}$ states of \ce{Ni} in \ce{\beta-NiUO4}.
While eigenstates $\nu_3$ to $\nu_5$ are fully occupied ($\lambda >0.98$) in both spin channels, $\lambda_1^\downarrow$ and $\lambda_2^\downarrow$ are significantly closer to zero (albeit still well above 0.1), indicating formally empty ($\widetilde{e_g}$) states.
The remaining eigenstates $\nu_3^\downarrow$ to $\nu_5^\downarrow$ are again fully occupied, which is consistent with the minority-spin $\widetilde{t_{2g}}$ states of a high-spin $d^8$ configuration.
This assignment is confirmed by the projected density of states (PDOS) shown in Figure~\ref{fig:pdos-isosurfaces}a, where it can be recognized that the \ce{Ni} contributions to the conduction bands are exclusively composed of $\widetilde{e_g}$-like $\nu_1^\downarrow$ and $\nu_2^\downarrow$ states (Figure~\ref{fig:pdos-isosurfaces}e). 
The occupancies of the $3d$ orbitals of \ce{Mn^{2+}} and \ce{Co^{2+}} in \ce{MnUO4} and \ce{CoUO4} show a qualitatively very similar trend (Table~S1), and are in line with the expected respective $d^5$ and $d^7$ high-spin configurations that were also found by \citeauthor{murphyTiltingDistortionRutileRelated2021}.
In all of the three materials, the $3d$ shells host formally empty yet significantly occupied $\widetilde{e_g}$-like eigenstates in the spin-minority manifold that are clearly distinguishable from other empty states by virtue of their elevated eigenvalues ($\lambda >0.1$), in the PDOSs (Figures~S1 and S2) and when visualized in real space (Figure~S8).
While it is generally desirable to correct all states contributing to the frontier orbitals, here, we chose to exclude the $\widetilde{e_g}$-like eigenstates from the Hubbard manifold because their elevated occupation eigenvalues (stemming from hybridization between A-$\widetilde{e_g}$ and ligand $p$ orbitals~\cite{sitSimple2011}) give rise to spurious energy and force contributions~\cite{solovyev2All31996,mackeOrbitalResolvedDFTMolecules2024}.
We emphasize that this can be seen as an \textit{ad hoc} solution to the unsatisfactory projectability of the $\widetilde{e_g}$ manifold when atomic-like projectors are employed.
Another solution (which we do not follow here) is to switch to a Wannier function projector basis, as practiced in Ref.~\citenum{murphyTiltingDistortionRutileRelated2021} (see the SI therein for a comparison of WF vs.~NAO occupation numbers).
A possible orbital-resolved manifold for the $3d$ shells therefore consists of the $\widetilde{t_{2g}}$ states, which are reasonably represented by the (ortho-)atomic projector orbitals and thus exhibit occupation eigenvalues close to either 0 or 1. 
This target subspace includes $\nu_3$–$\nu_5$ for \ce{\beta-NiUO4} (both spin channels), $\nu_3^\uparrow$–$\nu_5^\uparrow$ and $\nu_1^\downarrow$–$\nu_3^\downarrow$ for \ce{MnUO4}, and $\nu_3^\uparrow$–$\nu_5^\uparrow$, $\nu_1^\downarrow$, as well as $\nu_4^\downarrow$–$\nu_5^\downarrow$ for \ce{CoUO4}.
The assignment of the spin-majority eigenstates to either $\widetilde{t_{2g}}$ or $\widetilde{e_g}$ must rely on the eigenvectors, as the corresponding eigenvalues are all close to unity and thus do not allow for a distinction between the two manifolds.
Despite their numerical full occupation, the spin-majority $\widetilde{e_g}$-like eigenstates were also removed from the Hubbard manifold because fully occupied states are still affected by the Hubbard correction via the potential shift of $-U/2$ resulting from Eq.~\ref{eq:hubbard-potential-u}.
Moreover, an imbalanced correction of spin orbitals (i.e., correcting more spin-majority than spin-minority orbitals) would emulate a Hund $J$ term~\cite{himmetoglu_first-principles_2011,linscottRoleSpinCalculation2018}, which might be useful for addressing fractional spin errors, but a systematic treatment of such effects would go beyond the scope of this study.

For uranium, the PDOS indicates that most $5f$ contributions to the electronic structure are located in the conduction band (Figure~\ref{fig:pdos-isosurfaces}b); however, the projected occupations of some eigenstates are significantly larger than zero despite the formal $5f^0$ configuration of \ce{U^6+}.
Particularly, the high-energy eigenstates $\nu_5$, $\nu_6$ and $\nu_7$ exhibit eigenvalues above $0.2$ (in both spin channels), with $\lambda_7$ growing as large as $0.45$ in \ce{MnUO4} (Tables~\ref{tab:eigenvalues} and S1).
Given the continuous nature of the U-$5f$ occupation spectrum, the definition of an orbital-resolved Hubbard manifold is somewhat ambiguous and cannot be guided by symmetry considerations alone.
Based on their comparatively low occupation eigenvalues, we considered the four least-occupied eigenstates ($\nu_1$–$\nu_4$) to be sufficiently localized to be included in the Hubbard manifold, whereas $\nu_5$–$\nu_7$ were not targeted by Hubbard corrections to avoid overcorrection due to heavy ligand admixture.
To evaluate the robustness of our choice, we examined an alternative definition of the target subspace in which $\nu_5$ was also included. 
This variation led to only minor changes in the computed values of the $U$ parameters (Table~S2).

\begin{figure*}[ht!]
\centering
     \includegraphics[width=\textwidth]{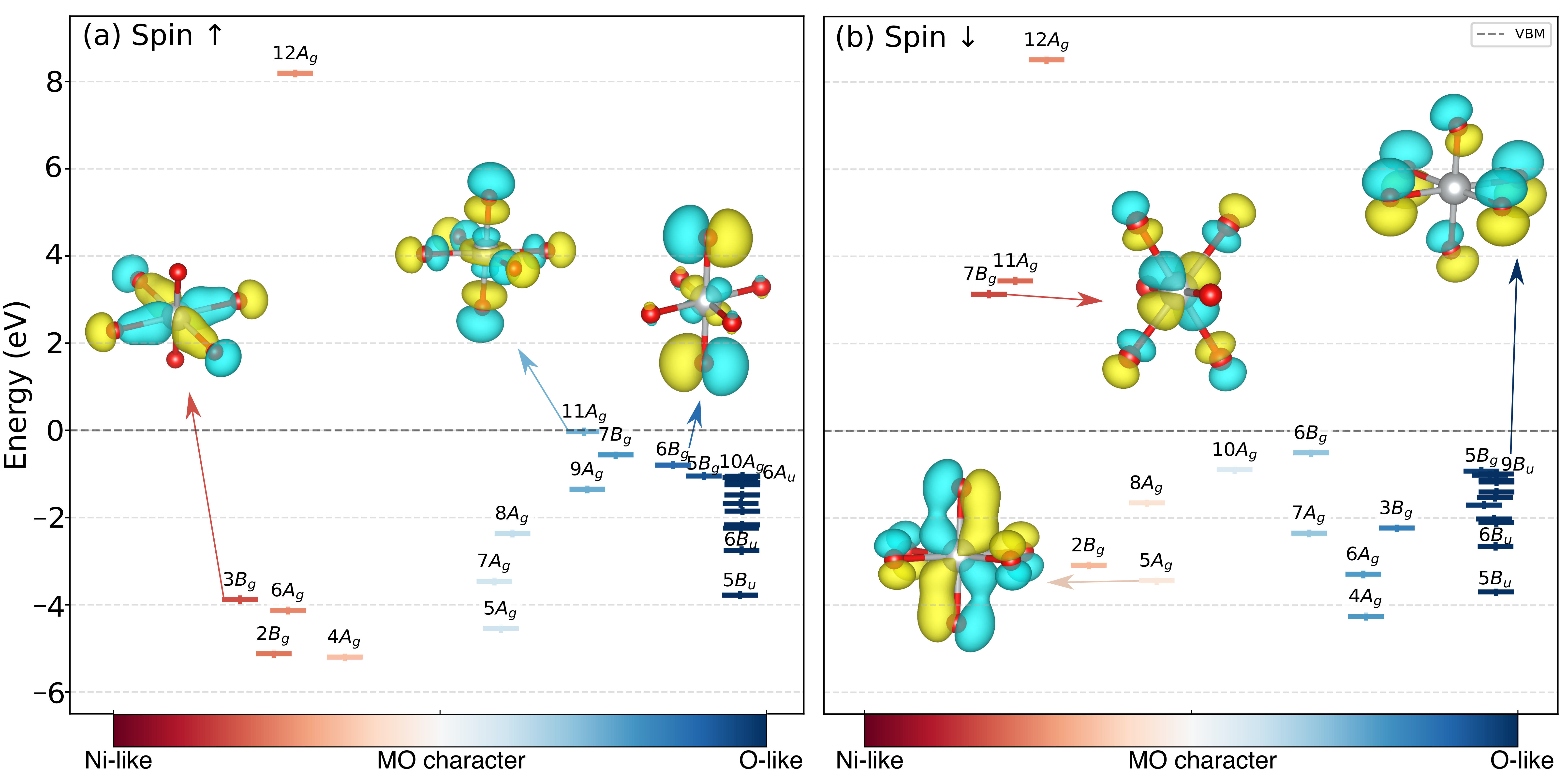}
      \caption{Molecular orbital (MO) diagram of an \ce{NiO6} octahedron in \ce{NiUO4} indicating the relative TM$-3d$ vs.~\ce{O}-$2p$ character of the MOs on the x-axis. The insets show the MOs corresponding to a few irreducible representations of interest. Data was generated by applying the linear combination of fragment orbitals method~\cite{mullerFragment2024} implemented in \textsc{LOBSTER}~\cite{maintzLOBSTER2016} to the ground state of the trial DFT+$U$(OAO) setup.}
       \label{fig:MO-diagram}
\end{figure*}

While partially filled $d$ and $f$ shells are typically the focus of Hubbard $U$ corrections, the possibility that \ce{O}-$2p$ states might also be localized~\cite{orhanFirstprinciplesHubbardHunds2020} (and thus require correction of SIEs) is often overlooked.
In the present ternary monouranates, the (distorted) trigonal planar coordination geometries of both \ce{O} sites, with bond angles between $105^{\circ}$ and $140^\circ$, suggest the presence of three $sp^2$ hybrid orbitals mediating the interactions between the oxygen atom and the metal centers, and one lone-pair orbital.
The $sp^2$ hybrid orbitals display significant $\sigma$-overlap with neighboring \ce{A}- and \ce{U}-site cations (for example, Figure~\ref{fig:pdos-isosurfaces}e shows a $\sigma^*_{d-p}$ orbital forming part of the lowest conduction bands), while the unhybridized $p_x$/$p_y$ orbitals (depending on whether the site is O1 or O2) interact only slightly with the neighboring sites via $\pi$ overlap.
The occupation eigenvalues of O-$2p$ are far from one, ranging from $\lambda \approx 0.73$ to 0.82 (Table~\ref{tab:eigenvalues} for \ce{\beta-NiUO4}, Table~S1 for \ce{MnUO4} and \ce{CoUO4}).
However, there is a noticeable gap between the numerical values of $\lambda_2$ ($\approx 0.76$) and $\lambda_3$ ($\approx 0.80$), except for the spin-up manifold of O2 where both values are similar. 
Careful analysis of the PDOS shows that energy ranges where A-$\widetilde{t_{2g}}$ states dominate the electronic structure (e.g., the red shaded area in Figure~\ref{fig:pdos-isosurfaces}a) also show large contributions due to $\nu_3$ of O-$2p$. 
By visualizing the eigenstates of the occupation matrix (Figure~S8), it becomes evident that $\nu_3$ corresponds to the nominally unhybridized $p_x$ ($p_y$) orbital of O1 (O2).
We argue that the orbital-resolved Hubbard manifold of \ce{O} should comprise only this eigenstate, which exhibits weaker hybridization and a more atomic-like character.
In contrast, the $sp^2$-hybridized states are improperly represented by (linear combinations of) $2p$ OAOs (Figure~S7), potentially rendering a correction of SIEs through Hubbard $U$ terms difficult.
This assessment is further backed by the orbital-energy diagram of the \ce{NiO6} octahedron depicted in Figure~\ref{fig:MO-diagram}, where it can be seen that a large number of symmetry-adapted linear combinations (SALCs) remain purely oxygenic.
These \textit{ungerade} states (e.g., $6A_u$, $9B_u$) account for most of the levels just below the valence-band maximum, are non-bonding to weakly antibonding (Figure~S9), and indeed show a strong atomic-like $p$ orbital character.
On the other hand, the \ce{O}-$2p$ contributions to the $11A_g$ frontier orbital, which is an antibonding $d-p\,\sigma^*$ molecular orbital (MO), display heavy admixture with \ce{Ni}-$d$ orbitals, highlighting the difficulty of representing these states with a conventional (ortho-)atomic projector basis.
For this reason $\nu_1$ and $\nu_2$ were not included in the orbital-resolved Hubbard manifold of \ce{O}-$2p$, whereas $\nu_3$ was.

\newpage

\subsection{Evaluation of the Hubbard \texorpdfstring{$U$}{U} parameters}
Having established orbital-resolved Hubbard manifolds for all frontier shells, we proceed to evaluate the respective Hubbard parameters from LR-cDFT and compare the outcome for different choices of Hubbard manifolds. 
In the traditional shell-averaged manifold, Hubbard corrections are applied to all A-$3d$, U-$5f$, and O-$2p$ orbitals. 
The OR-DFT+$U$ approach systematically refines this definition of the target subspace through the following sequence of setups with increasingly selective manifolds:
(1) The U-$5f$ subspace is restricted to eigenstates $\nu_1$–$\nu_4$, while shell-averaged $U$ corrections are retained for the A-site $3d$ and O-$2p$ states;
(2) as in (1), but with an orbital-resolved $U$ applied to A-site $\widetilde{t_{2g}}$ states;
(3) as in (2), but additionally limiting the O-$2p$ correction to the $p_x$ (O1) and $p_y$ (O2) states.

\begin{table}[hb!]
    \fontsize{8pt}{8pt}\selectfont
    \centering
    \setlength{\tabcolsep}{4pt}
    \renewcommand{\arraystretch}{1.6} 
    \caption{Self-consistent Hubbard $U$ parameters obtained from LR-cDFT for four setups (one shell-averaged plus three orbital-resolved) with differently defined Hubbard manifolds. The values in parentheses are those derived and applied by \citeauthor{murphyTiltingDistortionRutileRelated2021}.}
    \begin{tabular}{c c c c c}
        \arrayrulecolor{black}\hline
          System & \multicolumn{4}{c}{$U$ (eV)} \\  \hline
        \rowcolor{gray!10}[6pt] Shell-averaged
           & A-$3d$ & U-$5f$ & O1-$2p$ & O2-$2p$ \\ 
    
           \ce{\beta-NiUO4} & 7.03 (6.6) & 3.65 (2.6) & 7.56 & 7.97 \\ 
           \ce{MnUO4} & 7.34 (4.4) & 3.92 (2.7) & 7.45 & 7.82 \\ 
           \ce{CoUO4} & 6.19 (5.2) & 3.77 (2.7) & 7.54 & 7.94\\ 
    
        \rowcolor{gray!10}[6pt] Setup (1)
           & A-$3d$ & U-${5f}_{\nu_1-\nu_4}$ & O1-$2p$ & O2-$2p$ \\ 
    
           \ce{\beta-NiUO4} & 7.03 & 1.37 & 7.59 & 7.97 \\ 
           \ce{MnUO4} & 7.17 & 1.38 & 7.49 & 7.87 \\ 
           \ce{CoUO4} & 6.20 & 1.36 & 7.56 & 7.94\\ 
           
        \rowcolor{gray!10}[6pt] Setup (2)
           & A-${\widetilde{t_{2g}}}$ & U-${5f}_{\nu_1-\nu_4}$ & O1-$2p$ & O2-$2p$ \\ 
           
           \ce{\beta-NiUO4} & 7.40 & 1.33 & 7.50 & 7.78 \\ 
           \ce{MnUO4} & 1.27 & 0.93 & 7.40 & 7.74 \\ 
           \ce{CoUO4} & 2.33 & 1.22 & 7.47 & 7.77 \\ 
    
        \rowcolor{gray!10}[6pt] Setup (3)
           & A-${\widetilde{t_{2g}}}$ & U-${5f}_{\nu_1-\nu_4}$ & O1-$2p_x$ & O2-$2p_y$ \\ 
           
           \ce{\beta-NiUO4} & 7.68 & 1.26 & 4.14 & 4.13 \\
           \ce{MnUO4} & 1.46 & 1.01 & 4.25 & 4.18 \\
           \ce{CoUO4} & 2.56 & 1.15 & 4.11 & 4.14 \\        
        \arrayrulecolor{black}\hline       
    \end{tabular}
    \label{tab:parameters}
\end{table}

Table~\ref{tab:parameters} shows the $U$ values obtained for all choices of Hubbard manifolds and for each compound.
While still falling in typical ranges~\cite{orhanFirstprinciplesHubbardHunds2020,bastoneroFirstprinciplesHubbardParameters2025}, all shell-averaged $U$ values are comparatively large, and even exceed those reported by \citeauthor{murphyTiltingDistortionRutileRelated2021} by 1 to 3\,eV.
The reason for this discrepancy is likely rooted in the use of different Hubbard projector functions: while \citeauthor{murphyTiltingDistortionRutileRelated2021} employed nonorthogonalized atomic orbitals, here, OAO were used.

Moving forward to setup (1), excluding the most hybridized eigenstates $\nu_5-\nu_7$ from the Hubbard manifold of U-$5f$ prompts a strong and consistent reduction in the corresponding on-site interaction parameters of all three isomorphous compounds, from $U_{\mathrm{U\text{-}}5f}\approx 3.8\,$eV to $U_{\mathrm{U\text{-}}5f_{\nu_1-\nu_4}}\approx 1.4$\,eV. 
At the same time, the \ce{A}-$3d$ and \ce{O}-$2p$ manifolds barely respond to the exclusion of $\nu_5-\nu_7$, maintaining the same $U$ values as in the shell-averaged manifold.

A more drastic drop is observed in setup (2) for \ce{MnUO4} and \ce{CoUO4}, where substituting the shell-averaged $U_{\mathrm{A\text{-}3}d}$ correction by a $\widetilde{t_{2g}}$-specific one leads to $U$ parameters as low as \mbox{$\approx1.3\,$eV} (\ce{MnUO4}) and $\approx 2.3\,$eV (\ce{CoUO4}), respectively.
The 1\,eV difference between $U_{\mathrm{Mn\text{-}}\widetilde{t_{2g}}}$ and $U_{\mathrm{Co\text{-}}\widetilde{t_{2g}}}$ can be understood by recognizing that the $\widetilde{t_{2g}}$ subspace of \ce{Mn} contains three electrons [$\widetilde{t_{2g}}^3(\uparrow)$], whereas that of \ce{Co} holds five [$\widetilde{t_{2g}}^3(\uparrow)$ and $\widetilde{t_{2g}}^2(\downarrow)$].
In \ce{\beta-NiUO4}, however, the $\widetilde{t_{2g}}$ subspace is fully occupied [$\widetilde{t_{2g}}^3(\uparrow)$ and $\widetilde{t_{2g}}^3(\downarrow)$], which causes the manifold to be quite insensitive to perturbations, i.e., its occupations barely change during the course of the LR-cDFT calculations~\cite{kulikSystematicStudyFirstrow2010a}.
This results in a small numerical response that ultimately translates into a large orbital-resolved Hubbard parameter of $U_{\mathrm{Ni\text{-}}\widetilde{t_{2g}}}=7.4\,$eV.
Interestingly, the transition to orbitally-resolved A-$3d$ manifolds also affects the $U$ values of the other manifolds, particularly in \ce{MnUO4}, where $U_{\mathrm{U\text{-}}5f_{\nu_1-\nu_4}}$ drops by 0.4\,eV compared to setup (1), whereas $U_{\mathrm{O1\text{-}}2p}$ and $U_{\mathrm{O2\text{-}}2p}$ decrease only slightly by $\approx0.1\,$eV.
This demonstrates the importance of the $\widetilde{e_g}$ states for both intra-shell screening ($\widetilde{t_{2g}}\Leftrightarrow\widetilde{e_g}$) and inter-site interactions between the A-$3d$ shell and adjacent ligand $2p$ orbitals.

Finally, with setup (3) the $U$ values of oxygen also experience a strong reduction, as the Hubbard $U$ correction to O-$2p$ is restricted to the localized $p_x$ (O1) and $p_y$ (O2) orbitals, respectively. 
Remarkable is not only the drop in values by itself, from $\geq 7.4$ down to $\approx4.2$\,eV, but also the tightening of the spread (i.e., the difference between the largest and the smallest $U_{\mathrm{O\text{-}}2p}$ value), which amounts to $\approx0.5\,$eV in the previous setups but shrinks to only 0.1\,eV in setup (3). 
The fact that $U_{\mathrm{O\text{-}}2p_{x/y}}$ is essentially constant, regardless of the A-site cation or the crystallographic site, is a testimony to the almost nonbonding (lone-pair) nature of these orbitals.
Nevertheless, with $U$ values on the order of $4.2\,$eV, they severely contribute to deviations from PWL; and intriguingly do so much more than the A$-\widetilde{t_{2g}}$ and U$-5f_{\nu_1-\nu4}$ subspaces.

The strong reduction in $U$ parameters upon transition to orbital-resolved manifolds highlights an important feature of the linear-response formalism, namely that only screening channels outside the target subspace contribute to the effective interaction strength~\cite{vaugierHubbardHundExchange2012,linscottRoleSpinCalculation2018,mackeOrbitalResolvedDFTMolecules2024}.
When perturbing entire $5f$, $3d$, or $2p$ shells, important interactions (particularly intra-shell ones like $\widetilde{t_{2g}} \Leftrightarrow \widetilde{e_g}$) are prevented from screening the perturbation, which leads to large $U$ values due to small (apparent) responses. 
This problem is aggravated by the use of OAO projectors, which often misattribute ligand electrons (e.g., belonging to O-$sp^2$ hybrid orbitals) to metal orbitals, thus suppressing inter-shell screening pathways.
By instead targeting only the most localized frontier orbitals, one mitigates both issues: the relevant intra-shell screening channels remain active, and the misattribution of ligand electrons remains unpunished, as the $\widetilde{e_g}$ and \ce{U}-$5f_{\nu_5-\nu_7}$ manifolds are not included in the Hubbard manifold.
For the \ce{AUO4} systems investigated here, the orbitally-resolved interaction parameters are therefore expected to be more representative of the physical reality than their shell-averaged counterparts.

\subsection{Structural distortions}
    \label{subsec:calc-properties}
To assess the impact of differently defined Hubbard manifolds on crystal structure predictions, we evaluate the key distortion parameters $\theta$, $\delta_\mathrm{U}$ and $\delta_\mathrm{O}$ defined in Sec.~\ref{subsec:distortions-formulae}.
These distortions are highly sensitive to the underlying electronic structure (and vice versa), as illustrated by the PDOSs obtained using different DFT($+U$) schemes in Figures S3--S5 of the SI, and thus provide a meaningful benchmark for this purpose.
The PDOSs computed with the shell-averaged DFT+$U$ method [(b) panels] bear some similarity with those of DFT+$U$(WF) [(f) panels]; however, the latter predicts a valence band edge of predominant A-$3d$ character for \ce{MnUO4} and \ce{CoUO4}, whereas in the former, these band edges are strongly dominated by O-$2p$ states.
The gradual exclusion of the hybridized states from the Hubbard manifold [(c), (d), and (e) panels] leads to a PDOS more closely resembling that of PBEsol [(a) panels], with the \ce{A}-$3d$ dominated states located close to the valence band maximum.  
Regarding the band gaps, DFT+$U$(WF), shell-averaged DFT+$U$ method and hybrid functional (HSE06 \cite{heydHybrid2003}, provided here as a reference (Figure S6)) consistently yield gaps on the order of $\approx 2\,$eV, whereas smaller gaps $<1\,$eV are obtained for the OR-DFT+$U$ setups (2) and (3), except for \ce{MnUO4}, which is predicted to be metallic.
Moreover, the OR-DFT+$U$ setups (2) and (3) and HSE06 predict broader valence and conduction bands than DFT+$U$(WF) and shell-averaged DFT+$U$, whose bands are comparatively denser.
Given the lack of experimental data on the electronic structure of these materials, no definitive conclusion can be drawn regarding the spectral accuracy of the various methods.
Nevertheless, although DFT+$U$(WF) qualitatively gives the best match to the HSE06 results, OR-DFT+$U$ setups (2) and (3) show better agreement in terms of band widths.
These subtle similarities and differences are taken into account when discussing accuracy of the different methods in predicting the structural distortions.  

\begin{figure*}[ht!]
    \centering    
    \includegraphics[width=1.0\textwidth]{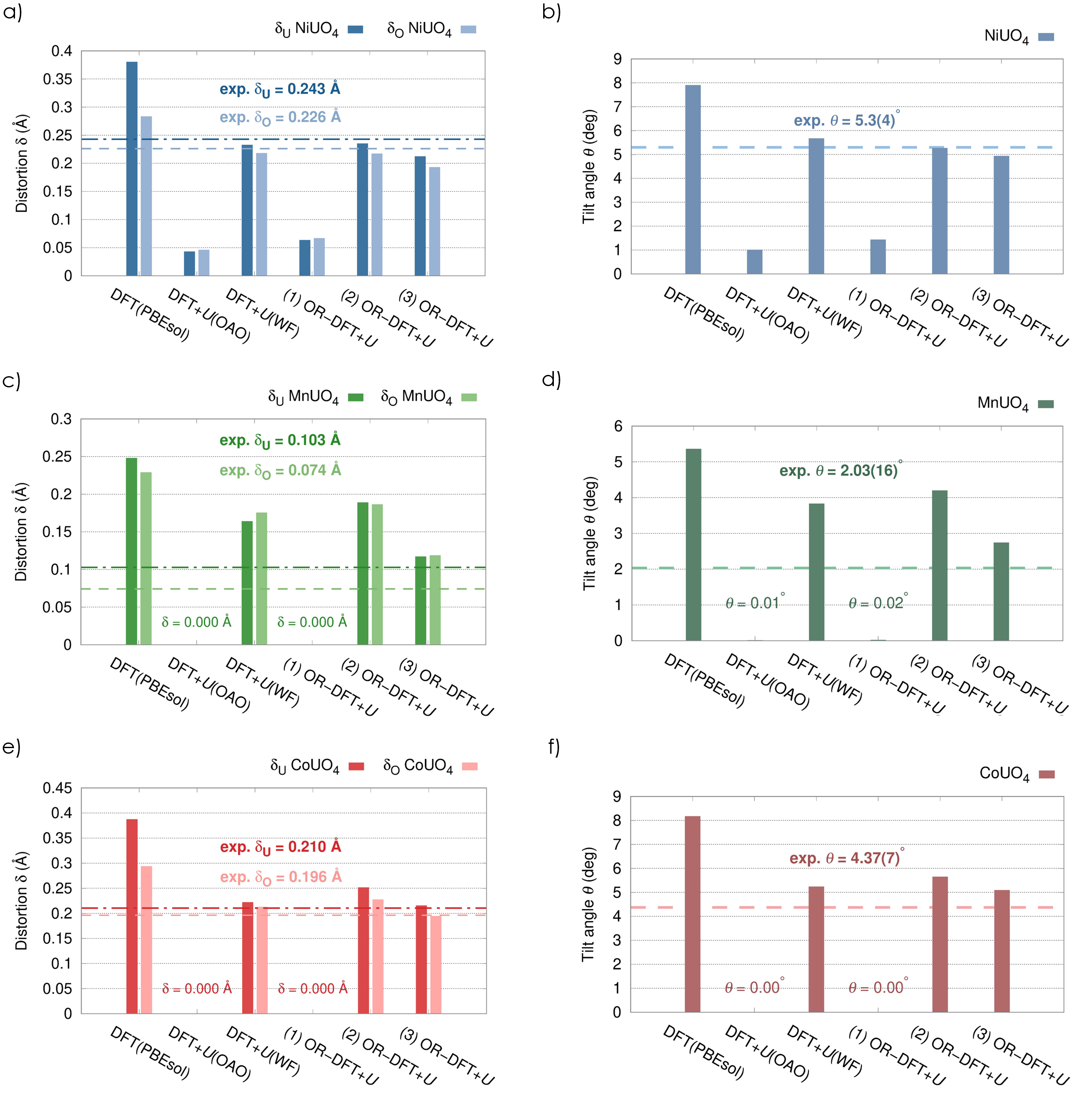}
    \caption{Off-center displacements of uranium, $\delta_\mathrm{U}$, and O2 atoms, $\delta_\mathrm{O}$ (a,c,e), and octahedral tilt angle $\theta$ in (b,d,f) as obtained from bare PBEsol and DFT$+U$ calculations using differently defined Hubbard manifolds (see text and Table~\ref{tab:parameters}). The DFT+U(WF) data points were taken from Ref.~\citenum{murphyTiltingDistortionRutileRelated2021}.}
    \label{fig:distortions}
\end{figure*}

Figure~\ref{fig:distortions} presents the results for uncorrected DFT (PBEsol), the shell-averaged manifold (shell-averaged DFT$+U$) and the three orbital-resolved Hubbard manifolds defined above, applied using the parameters reported in Table~\ref{tab:parameters}.
Also shown for comparison is the DFT$+U$(WF) data taken from in Ref.~\citenum{murphyTiltingDistortionRutileRelated2021} (details of these calculations are provided in Section IA of the SI).
Importantly, the projector basis (OAO) is identical for the shell-averaged calculations and the orbitally-resolved ones, so that differences between these approaches arise solely from the specification of the target subspace rather than from the projectors.

The uncorrected PBEsol functional significantly overestimates the lattice distortions in all three compounds.
The most pronounced case is \ce{MnUO4} (Figure~\ref{fig:distortions}c/d), where the predicted values of $\delta_\mathrm{U}$ and $\theta$ massively exceed the experimental estimates~\cite{murphyTiltingDistortionRutileRelated2021}, by 241\% and 264\%, respectively.
For \ce{\beta-NiUO4} and \ce{CoUO4} (Figure~\ref{fig:distortions}a/b and e/f, respectively), the deviations are slightly less dramatic but remain significant; for instance, $\theta$ is off by nearly 90\% in \ce{CoUO4}.

If these overestimations stem from SIEs (and the associated spurious over-delocalization of charge), Hubbard $U$ corrections should rectify or at least mitigate them, provided that the Hubbard manifold is defined such that (only) localized frontier states are targeted by the correction~\cite{zhao_global_2016,mackeOrbitalResolvedDFTMolecules2024}.
This is seemingly not the case for shell-averaged DFT+$U$, which suppresses structural distortions across all compounds.
While in \ce{\beta-NiUO4}, $\delta_\mathrm{U}$, $\delta_\mathrm{O}$, and $\theta$ are severely underestimated by up to 500\%, \ce{MnUO4} and \ce{CoUO4} are even entirely transformed into the higher-symmetry $Cmmm$ configurations, as all of of their distortion parameters drop to zero (barring numerical noise).
The DFT+$U$(WF) scheme represents a notable improvement with respect to both uncorrected DFT and shell-averaged DFT+$U$, as the Wannier function basis adequately represents frontier orbitals that preserve spatial localization while minimizing overlap with ligand states.
With the improved WF Hubbard projectors, the distortion parameters calculated for \ce{\beta-NiUO4} and \ce{CoUO4} are found in good agreement with the experiment, deviating by less than 20\%.
Only for \ce{MnUO4} the accuracy is less satisfactory, as $\delta_\mathrm{U}$, $\theta$ and $\delta_\mathrm{O}$ remain overestimated by 59\%, 89\% and 136\%, respectively.
The reasons for this inconsistency will be discussed later.

OR-DFT+$U$ with setup (1) offers no improvement over shell-averaged DFT$+U$, as the structural distortions remain suppressed almost entirely across all compounds.
This is in spite of the considerable difference between the numerical values of $U_{\mathrm{U\text{-}5}f_{\nu_1-\nu_4}}$ and $U_{\mathrm{U\text{-}5}f}$.
However, the Hubbard corrections to U$-5f$ mainly affect states in the conduction band, which clearly do not influence the ground-state ionic structures.
This result supports previous findings indicating that the symmetry of ternary monouranates is mainly controlled by the \ce{A}-site cations~\cite{baurCoReO4NewRutiletype1992,murphyTiltingDistortionRutileRelated2021}.

Hence, expanding the orbital resolution to the A-sites' $\widetilde{t_{2g}}$ orbitals within setup~(2) markedly improves the accuracy of predictions for \ce{\beta-NiUO4} and \ce{CoUO4}, with $\delta_\mathrm{U}$, $\delta_\mathrm{O}$ and $\theta$ closely matching both experimental values and the results of DFT+$U$(WF).
However, and similar to DFT$+U$(WF), setup (2) still overestimates the distortions in \ce{MnUO4}.
It is worth noting that \ce{Mn^{2+}} ions are comparatively large, with an ionic radius of 0.83\,\AA~ that exceeds the radii of both \ce{Co^{2+}} (0.745\,\AA) and \ce{Ni^{2+}} (0.69\,\AA)~\cite{shannonRevisedEffectiveIonic1976}.
Consequently, \ce{Mn^{2+}} shows a stronger tendency to hybridize with neighboring \ce{O}-$2p$ states that is also reflected in the more pronounced variations of the Hubbard $U$ values (Table~\ref{tab:parameters}).
In view of this, it is no surprise that setup (3), where the Hubbard manifold of \ce{O} is restricted to the localized $p_x$/$p_y$ orbitals (in addition to orbitally-resolved corrections to \ce{U}-$5f$ and \ce{A}-$3d$), cures the systematic overestimation of distortion parameters in \ce{MnUO4}.
Fortunately, the orbital-resolved treatment of \ce{O}-$2p$ affects the already good predictions of the other two compounds to a much lesser degree: minor improvements result for \ce{CoUO4}, where $\delta_\mathrm{U}$ and $\delta_\mathrm{O}$ match the experimental value almost exactly, whereas the accuracy is minimally deteriorated for \ce{\beta-NiUO4}.
Therefore, setup (3) provides the most consistent overall accuracy, delivering very good predictions for all distortion parameters and across all compounds.
Its dramatic edge over setup (2) and DFT$+U$(WF) in \ce{MnUO4} shows the importance of avoiding the correction of strongly hybridized states, here consisting in the bonding $\sigma$-states that form between \ce{A}-$\widetilde{e_g}$ and the \ce{O}-$sp^2$ hybrid orbitals.
More in general, the improvements achieved due to setups (2) and (3) demonstrate that a physically meaningful use of Hubbard $U$ corrections requires careful disentanglement of localized and delocalized states.
This is particularly crucial for systems where orbital hybridization plays an important role.
Additional relative energies of the \textit{Cmmm} and \textit{Ibmm} phases (Table S4) corroborate the structural analysis: setups (2) and (3) yield the correct ground-state ordering for \ce{\beta-NiUO4} and \ce{CoUO4}, and the residual \ce{MnUO4} deviation vanishes after removing the Hubbard contribution, consistent with Ref.~\citenum{murphyTiltingDistortionRutileRelated2021}.

\subsection{Projector mismatch as a source of spurious Hubbard forces} 
The profound impact of the Hubbard manifold on the accuracy of structural predictions for \ce{AUO4} compounds demands clarification.
While it is generally known that the ionic ground state of DFT$+U$ can differ significantly from that of the bare functional, it is not understood why and under which circumstances DFT$+U$ leads to over-symmetrization of atomistic structures, as observed here and reported elsewhere for diverse TM oxides~\cite{murphyTiltingDistortionRutileRelated2021,doDelocalizedPolaronBursteinMoss2023,gebreyesusUnderstandingRoleHubbard2023,liuAnomalousReversalStability2025,cartaExplicitDemonstrationEquivalence2025}.
In the following, we try to identify the root cause of these deviations and demonstrate why certain Hubbard projector functions or target manifolds perform better than others.

It follows from the Hellmann-Feynmann theorem that the Hubbard energy functional (c.f.~Equation~\ref{eq:edftu}) contributes an additional term $\mathbf{F}_U^I$ to the total force acting on an ion $I$~\cite{himmetoglu_hubbard-corrected_2014,timrovPulayForcesDensityfunctional2020}:
\begin{equation}
    \label{eq:force-hub}
    \mathbf{F}_U^I = - \frac{\partial E_U}{\partial \mathbf{R}^I} = - \sum_{\mathbf{k},v,\sigma} \left\langle \psi_{\mathbf{k},v}^\sigma \left| \frac{\partial V_{U}^\sigma}{\partial \mathbf{R}^I} \right| \psi_{\mathbf{k},v}^\sigma \right\rangle \, ,
\end{equation}
where $\mathbf{R}_I$ denotes the position of the $I$th atom. 
Eq.~\ref{eq:force-hub} implies that a finite Hubbard force arises whenever the derivative $\frac{\partial {V}_{U}^\sigma}{\partial \mathbf{R}^I}$ is non-zero.
Recalling Eq.~\ref{eq:hubbard-potential-u} and assuming a negligible dependence of $U$ on the position, this condition is fulfilled if a displacement of ion $I$ ($\partial\mathbf{R}_I$) modifies the occupation of an eigenstate ($\partial\lambda^{I\sigma}_i$), for instance due to an increase or decrease in its overlap with occupied KS states.
Note that $\frac{\partial {V}_{U}^\sigma}{\partial \mathbf{R}^I}$ can also be non-zero if the projectors themselves vary, e.g., due to re-orthogonalization of OAO projectors.
In the following, however, we neglect these so-called Pulay forces~\cite{timrovPulayForcesDensityfunctional2020} and focus instead on the electronic ``density response''~\cite{florisVibrationalPropertiesMnO2011}.
Because of the form of the Hubbard energy functional, eigenstates with occupation $\lambda_i^{I\sigma} < 0.5$ will experience a force that tends to further reduce their occupation (driving $\lambda_i^{I\sigma} \rightarrow 0$) by lowering their overlap with occupied KS wavefunctions. Conversely, eigenstates with $\lambda_i^{I\sigma} > 0.5$ will tend to increase their occupation.
The central question is: why would an eigenstate exhibit an occupation far from both zero and one to begin with?
This consideration allows us to distinguish two fundamentally different scenarios under which non-zero Hubbard forces arise:

In scenario (i), $\frac{\partial \lambda_i^{I\sigma}}{\partial \mathbf{R}^I}\neq 0$ because the occupation of an eigenstate deviates from zero or one due to violations of the PWL condition.
Hence, the resulting forces act to counter the artificial delocalization of states caused by SIEs.
In scenario (ii), $\frac{\partial \lambda_i^{I\sigma}}{\partial \mathbf{R}^I}\neq 0$ because the (localized) Hubbard projector functions do not accurately represent the true electron distribution in the system and produce occupation numbers that lie between zero and one, independently of the presence of SIEs.
Here, the resulting forces bias the charge density distribution $\rho(\mathbf{r})$ (and therefore also the ionic structure) towards alignment with the Hubbard projector functions.
Such forces are artificial and unphysical, since there is no fundamental reason why the electron density of a system should conform to that of an arbitrarily defined set of projector functions.
Popular atomic-like projector functions, for example (including NAO, OAO, PAW, and LMTO projectors), are constructed based on radial Schrödinger equations for isolated atoms in their neutral charge state.
However, the isolated-atom model may not be a reliable approximation for real compounds.
This is especially the case for materials with strong covalent bonding (where the quantum numbers $n$, $l$ and $m$ can lose their physical meaning), for structures with highly distorted bond angles, or when the radius of a charged ion deviates significantly from that of the neutral atom.

\begin{figure*}[hb!]
    \centering
    \includegraphics[width=3.33in]{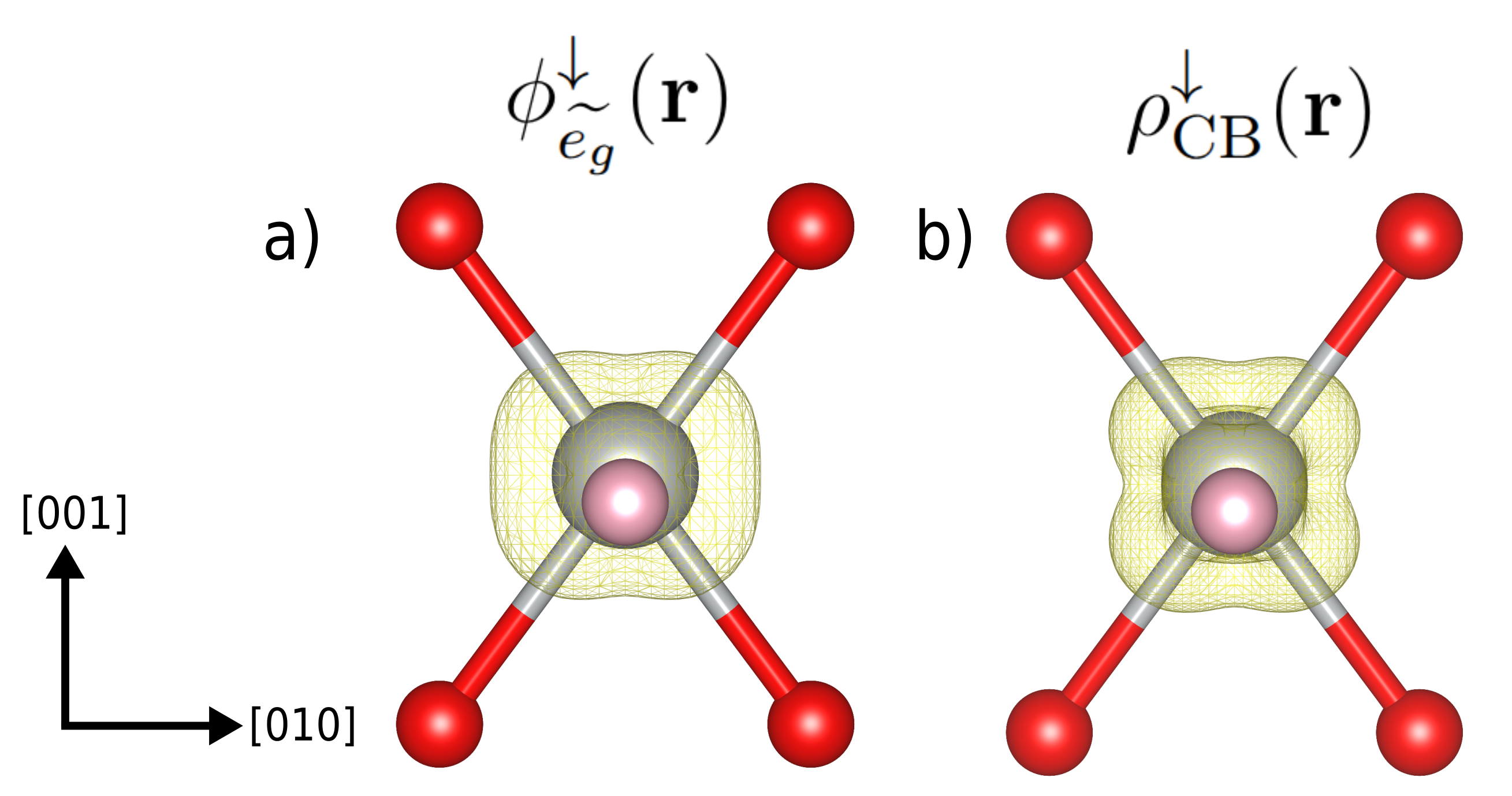}
    \caption{Comparison for \ce{\beta-NiUO4} between the spatial extension of (a) the squared modulus of the Hubbard projector orbitals, $|\phi_{\widetilde{e_g}}^\downarrow(\mathbf{r})|^2 \equiv  |\phi_1^\downarrow(\mathbf{r}) + \phi_2^\downarrow(\mathbf{r})|^2 = |\sum_{m=1}^5\left(\sum_{i=1}^2\nu_{mi}^\downarrow\right)\varphi_m(\mathbf{r})|^2$, representing the formally unoccupied $\widetilde{e_g}$-like orbitals, and (b) the (pseudo)-charge density due to the KS wavefunctions of the lowest conduction bands in the spin $\downarrow$ channel in the range from 1.8 to 4.2~eV w.r.t. valence band maximum, $\rho^\downarrow_\text{CB}(\mathbf{r}) = \sum_{v\in \text{CB}}\sum_{\mathbf{k}}  \left|\psi_{\mathbf{k},v}^\downarrow(\mathbf{r})\right|^2$ [see Figure~\ref{fig:pdos-isosurfaces}(a)]. It is the mismatch between projectors and KS wavefunctions that gives rise to the fractional occupation numbers $\lambda_1^\downarrow$ and $\lambda_2^\downarrow$. The color code is the same as in Figure~\ref{fig:structure}.}  
    \label{fig:projectors-vs-ks}
\end{figure*}
Figure~\ref{fig:projectors-vs-ks} illustrates this issue by showing the mismatch between the unoccupied (spin-down) $\widetilde{e_g}$-like KS orbitals around the \ce{Ni^{2+}} ion in \ce{\beta-NiUO4} and the diagonal projector orbitals $\phi_{\widetilde{e_g}}^\downarrow(\mathbf{r})$ corresponding to eigenstates $\nu_1^\downarrow$ and $\nu_2^\downarrow$ of the occupation matrix.
These eigenstates, which are linear combinations of the OAO projectors (Figures~S7 and S8), are indeed fractionally occupied ($\lambda_1^\downarrow \approx 0.16$, $\lambda_2^\downarrow \approx 0.22$, c.f. Table~\ref{tab:eigenvalues}).
Therefore, applying Hubbard $U$ corrections to this manifold induces an unphysical symmetrization of the structure, as the minimum of this DFT$+U$ ground state shifts to a configuration where the overlap of the occupied KS states with $\nu_1$ and $\nu_2$ is minimized.
Note that such artifacts often remain negligible in highly symmetric systems, including the well-investigated solids \ce{FeO} and \ce{NiO}.
These systems exhibit coordination polyhedra with ideal $O_h$ site symmetries, where the angular shape of atomic projectors (e.g., $90^\circ$ or $180^\circ$ for $d$ orbitals) coincides with the metal-ligand bond axes.

The observation of an artificial force contribution provides a rationale for the improvements observed when employing either WF projectors or OR Hubbard manifolds.
In the former case, it is the projector functions themselves that adapt to the charge density, thus resulting in a less fractional eigenvalue spectrum that yields no spurious force contributions.
Conversely, in the latter case, the inadequate eigenstates are simply excluded from the correction by setting $U_i=0$ for all respective $i$.
Note that spurious Pulay forces due to projector mismatch can also entail ``under-symmetrization'', as demonstrated by the improvements in the structural predictions for \ce{MnUO4} upon switching from OR manifold (2) to manifold (3). 
In manifold (2), the correction targets three $sp^2$ hybrid orbitals (with bond angles of approximately $120^\circ$) plus an unhybridized non-bonding state using three (ortho-)atomic $2p$ projectors ($p_x$, $p_y$, $p_z$).
While the non-bonding state (eigenstate $\nu_3$) is well represented by one of the projectors, it is evident that the remaining two projectors (or any linear combination thereof) cannot reproduce the spatial extent of the three $sp^2$ hybrid orbitals.
Consequently, correcting eigenstates $\nu_1$ and $\nu_2$ of the \ce{O}-$2p$ shell drives the trigonal planar geometry toward an atomic-like one, in which the bond angles are closer to $90^\circ$.
In the \ce{AUO4} structures considered here, such distortions increase the deviation from the more symmetric $Cmmm$ geometry, most significantly in \ce{MnUO4}, where interactions between the \ce{A}-site and neighboring \ce{O} atoms are strongest due to the relatively large radius of \ce{Mn^{2+}}.

\section{Concluding remarks}
    \label{sec:conclusions}
Using an orbital-resolved DFT$+U$ scheme~\cite{mackeOrbitalResolvedDFTMolecules2024}, we have demonstrated how the choice of the Hubbard manifold affects the ground-state electronic and structural properties of \ce{MnUO4}, \ce{CoUO4} and \ce{$\beta$-NiUO4}.
With (ortho-)atomic projector functions, the characteristic distortions of these monouranates are accurately reproduced only when the Hubbard manifold is restricted to the most localized subset of orbitals within the \ce{A}-$3d$, \ce{U}-$5f$, and \ce{O}-$2p$ shells.
In contrast, applying Hubbard $U$ corrections to more delocalized states within these shells markedly reduces the accuracy.
An exception is found for the \ce{U}-$5f$ shell, where the structural distortions are insensitive to the choice of Hubbard manifold, although changes in the electronic structure are still observed (see Figures~S3--S5).
The excessive sensitivity of the ionic structure to the choice of the Hubbard manifold stems from an implicit, artificial contribution in the Hubbard force expression, which aligns the computed charge density $\rho(\mathbf{r})$ (and thus, the ionic positions) with the diagonalized (ortho-)atomic projectors $\phi_i^{I\sigma}(\mathbf{r})$.

Two strategies can help mitigate these artifacts: either one employs Hubbard projector functions that reflect the local bonding environment more faithfully---such as Wannier functions (as practiced in Ref.~\citenum{murphyTiltingDistortionRutileRelated2021})---or one excludes strongly hybridized states (where the mismatch between $\rho(\mathbf{r})$ and $\phi_i^{I\sigma}(\mathbf{r})$ is large) from the Hubbard manifold using OR-DFT$+U$, as done here.
The success of both approaches, DFT+$U$(WF) and OR-DFT+$U$, hinges on an appropriate choice of the Hubbard manifold (in fact, this applies to all DFT$+U$ approaches, including the shell-averaged one).
Within the DFT$+U$(WF) framework, this involves the selection of energy windows and sometimes also the disentanglement of bands; for OR-DFT$+U$ with atomic-like Hubbard projectors, one must carefully determine which eigenstates of the occupation matrix exhibit true on-site character, e.g., by analyzing the eigenvalue spectrum.
These steps typically require chemical intuition and some degree of trial-and-error, although recent progress on the automation of the Wannierzation step~\cite{qiaoAutomatedMixingMaximally2023,qiaoProjectabilityDisentanglementAccurate2023} has significantly reduced the effort associated with using the DFT$+U$(WF) formalism.
DFT+$U$(WF) might outperform OR-DFT$+U$ in structures where the mismatch between the true charge density and the shape of atomic-like projectors is extreme, or where Hubbard corrections need to be applied to electronic states localized between atoms (cf. inter-site terms of Refs.~\citenum{leiria_campo_jr_extended_2010,cartaExplicitDemonstrationEquivalence2025}).
This can be important when atomic-like projector functions cannot adequately describe a localized manifold, for instance highly localized $sp^3$ hybrid orbitals.
A WF-based approach could still work in such cases because Wannier functions do not make any assumptions regarding the spatial extension of the localized states.
On the other hand, owing to its conceptual simplicity, OR-DFT$+U$ can leverage existing routines for the calculation of forces and stresses, thereby enabling structural relaxations and \textit{ab initio} molecular dynamics, which are often not feasible or not yet available within existing DFT$+U$(WF) implementations.
Furthermore, due to a more realistic representation of intra-shell screening, OR Hubbard manifolds often acquire lower first-principle $U$ values~\cite{mackeOrbitalResolvedDFTMolecules2024}, making the correction even more surgical.
Importantly, however, the case of \ce{\beta-NiUO4} demonstrates that the improvements of OR-DFT$+U$ with respect to shell-averaged schemes are mainly rooted in the exclusion of hybridized states from the Hubbard manifold, and only to a lesser extent (if any) in the reduction of the $U$ values.

In closing, the results of the this work suggest that studying the ground state of structurally complex compounds with distorted polyhedra, mixed valence, and hybridized electronic states requires a chemically informed and orbitally resolved framework, which both OR-DFT$+U$ and DFT$+U$(WF) provide.
A priority for future methodological advancements should be the development of nonempirical, quantitative protocols for identifying Hubbard manifolds (for a given class of projectors) with minimal user input and without reliance on chemical intuition.

\begin{acknowledgement}

E.M acknowledges funding by MaX ``Materials Design at the Exascale'', a Center of Excellence co-funded by the European High Performance Computing Joint Undertaking (JU) and participating countries under grant agreement No. 101093374.
I.T. acknowledges support from the Swiss National Science Foundation (SNSF), grant No.~200021-227641 and No.~200021-236507, and support from the NCCR MARVEL, a National Centre of Competence in Research, funded by SNSF. The authors gratefully acknowledge computing time on the supercomputer JURECA \citep{JURECA} at Forschungszentrum Jülich under grant No. cjiek61.

\end{acknowledgement}

\begin{suppinfo}

Occupation eigenvalues for \ce{MnUO4} and \ce{CoUO4}, projected density of states (PDOS) plots for all Hubbard manifolds and compounds presented, visualization of the OAO projector orbitals and the eigenstates of \ce{\beta-NiUO4}, and a detailed description of the tool used to visualize Hubbard projector functions in real space.

\end{suppinfo}

\bibliography{bibliography}

@article{jainComputational2016,
  title = {Computational Predictions of Energy Materials Using Density Functional Theory},
  author = {Jain, Anubhav and Shin, Yongwoo and Persson, Kristin A.},
  year = {2016},
  month = jan,
  journal = {Nat. Rev. Mater.},
  volume = {1},
  number = {1},
  pages = {15004},
  issn = {2058-8437},
  doi = {10.1038/natrevmats.2015.4},
  urldate = {2025-08-13},
  langid = {english}
}

@article{heydHybrid2003,
  title = {Hybrid Functionals Based on a Screened {{Coulomb}} Potential},
  author = {Heyd, Jochen and Scuseria, Gustavo E. and Ernzerhof, Matthias},
  year = 2003,
  month = may,
  journal = {The Journal of Chemical Physics},
  volume = {118},
  number = {18},
  pages = {8207--8215},
  issn = {0021-9606, 1089-7690},
  doi = {10.1063/1.1564060},
  urldate = {2023-09-13},
  langid = {english}
}

@article{maintzLOBSTER2016,
  title = {{{LOBSTER}}: {{A}} Tool to Extract Chemical Bonding from Plane-wave Based {{DFT}}},
  shorttitle = {{{LOBSTER}}},
  author = {Maintz, Stefan and Deringer, Volker L. and Tchougr{\'e}eff, Andrei L. and Dronskowski, Richard},
  year = 2016,
  month = apr,
  journal = {Journal of Computational Chemistry},
  volume = {37},
  number = {11},
  pages = {1030--1035},
  issn = {0192-8651, 1096-987X},
  doi = {10.1002/jcc.24300},
  urldate = {2025-10-25},
  copyright = {http://creativecommons.org/licenses/by-nc-nd/4.0/},
  langid = {english}
}

@article{mullerFragment2024,
  title = {Fragment {{Orbitals Extracted}} from {{First-Principles Plane-Wave Calculations}}},
  author = {M{\"u}ller, Peter C. and Schmit, Nathalie and Sann, Leander and Steinberg, Simon and Dronskowski, Richard},
  year = 2024,
  month = oct,
  journal = {Inorganic Chemistry},
  volume = {63},
  number = {43},
  pages = {20161--20172},
  issn = {0020-1669, 1520-510X},
  doi = {10.1021/acs.inorgchem.4c01024},
  urldate = {2025-10-25},
  copyright = {https://doi.org/10.15223/policy-029},
  langid = {english}
}

@article{JURECA,
author = {{J\"{u}lich Supercomputing Centre}},
title = {{JURECA: Data Centric and Booster Modules implementing the Modular Supercomputing Architecture at J\"{u}lich Supercomputing Centre}},
journal = {JLSRF},
number = {A182},
volume = {7},
doi = {10.17815/jlsrf-7-182},
url = {http://dx.doi.org/10.17815/jlsrf-7-182},
year = {2021}
}

@inbook{murphySolidState2019,
  title = {The {{Solid-State Chemistry}} of {{AUO}}{\textsubscript{4}} {{Ternary Uranium Oxides}}: {{A Review}}},
  shorttitle = {The {{Solid-State Chemistry}} of {{AUO}}{\textsubscript{4}} {{Ternary Uranium Oxides}}},
  booktitle = {Complex {{Oxides}}},
  author = {Murphy, Gabriel L. and Zhang, Zhaoming and Kennedy, Brendan J.},
  year = {2019},
  month = may,
  pages = {103--130},
  publisher = {WORLD SCIENTIFIC},
  doi = {10.1142/9789813278585_0004},
  urldate = {2025-08-13},
  collaborator = {Vogt, Thomas and Buttrey, Douglas J},
  isbn = {978-981-327-857-8 978-981-327-858-5},
  langid = {english}
}

@article{Romero2014,
author = {Blanca Romero, Ariadna and Kowalski, Piotr M. and Beridze, George and Schlenz, Hartmut and Bosbach, Dirk},
title = {Performance of DFT+U method for prediction of structural and thermodynamic parameters of monazite-type ceramics},
journal = {J. Comput. Chem.},
volume = {35},
number = {18},
pages = {1339-1346},
doi = {https://doi.org/10.1002/jcc.23618},
url = {https://onlinelibrary.wiley.com/doi/abs/10.1002/jcc.23618},
eprint = {https://onlinelibrary.wiley.com/doi/pdf/10.1002/jcc.23618},
year = {2014}
}

@article{anderssonDensityFunctionalTheory2013,
  title = {Density {{Functional Theory Calculations}} of {{UO}}{\textsubscript{2}} {{Oxidation}}: {{Evolution}} of {{UO}}{\textsubscript{2+{\emph{x}} }}, {{U}}{\textsubscript{4}}{{O}}{\textsubscript{9--{\emph{y}} }}, {{U}}{\textsubscript{3}}{{O}}{\textsubscript{7}}, and {{U}}{\textsubscript{3}}{{O}}{\textsubscript{8}}},
  shorttitle = {Density {{Functional Theory Calculations}} of {{UO}}{\textsubscript{2}} {{Oxidation}}},
  author = {Andersson, D. A. and Baldinozzi, G. and Desgranges, L. and Conradson, D. R. and Conradson, S. D.},
  year = {2013},
  month = mar,
  journal = {Inorg. Chem.},
  volume = {52},
  number = {5},
  pages = {2769--2778},
  issn = {0020-1669, 1520-510X},
  doi = {10.1021/ic400118p},
  urldate = {2025-05-19},
  langid = {english}
}

@article{baurCoReO4NewRutiletype1992,
  title = {{{CoReO4}}, a New Rutile-Type Derivative with Ordering of Two Cations},
  author = {Baur, Werner H. and Joswig, Werner and Pieper, Gerhard and Kassner, Dethard},
  year = {1992},
  month = jul,
  journal = {J. Solid State Chem.},
  volume = {99},
  number = {1},
  pages = {207--211},
  issn = {00224596},
  doi = {10.1016/0022-4596(92)90307-H},
  urldate = {2025-05-19},
  copyright = {https://www.elsevier.com/tdm/userlicense/1.0/},
  langid = {english}
}

@article{beridzeBenchmarkingDFTMethod2014,
  title = {Benchmarking the {{DFT}}+ {{{\emph{U}}}} {{Method}} for {{Thermochemical Calculations}} of {{Uranium Molecular Compounds}} and {{Solids}}},
  author = {Beridze, George and Kowalski, Piotr M.},
  year = {2014},
  month = dec,
  journal = {J. Phys. Chem. A},
  volume = {118},
  number = {50},
  pages = {11797--11810},
  issn = {1089-5639, 1520-5215},
  doi = {10.1021/jp5101126},
  urldate = {2025-05-19},
  langid = {english}
}

@article{bosbachResearchSafeManagement2020,
  title = {Research for the {{Safe Management}} of {{Nuclear Waste}} at {{Forschungszentrum J{\"u}lich}}: {{Materials Chemistry}} and {{Solid Solution Aspects}}},
  shorttitle = {Research for the {{Safe Management}} of {{Nuclear Waste}} at {{Forschungszentrum J{\"u}lich}}},
  author = {Bosbach, Dirk and Brandt, Felix and Bukaemskiy, Andrey and Deissmann, Guido and Kegler, Philip and Klinkenberg, Martina and Kowalski, Piotr M. and Modolo, Giuseppe and Niemeyer, Irmgard and Neumeier, Stefan and Vinograd, Victor},
  year = {2020},
  month = jun,
  journal = {Adv. Eng. Mater.},
  volume = {22},
  number = {6},
  pages = {1901417},
  issn = {1438-1656, 1527-2648},
  doi = {10.1002/adem.201901417},
  urldate = {2025-05-19},
  langid = {english}
}

@article{chenDFTStudyUranium2022,
  title = {{{DFT}}+{{{\emph{U}}}} {{Study}} of {{Uranium Dioxide}} and {{Plutonium Dioxide}} with {{Occupation Matrix Control}}},
  author = {Chen, Jia-Li and Kaltsoyannis, Nikolas},
  year = {2022},
  month = jul,
  journal = {J. Phys. Chem. C},
  volume = {126},
  number = {27},
  pages = {11426--11435},
  issn = {1932-7447, 1932-7455},
  doi = {10.1021/acs.jpcc.2c03804},
  urldate = {2025-05-19},
  copyright = {https://creativecommons.org/licenses/by/4.0/},
  langid = {english}
}

@article{chroneosNuclearWasteformMaterials2013,
  title = {Nuclear Wasteform Materials: {{Atomistic}} Simulation Case Studies},
  shorttitle = {Nuclear Wasteform Materials},
  author = {Chroneos, A. and Rushton, M.J.D. and Jiang, C. and Tsoukalas, L.H.},
  year = {2013},
  month = oct,
  journal = {J. Nucl. Mater.},
  volume = {441},
  number = {1-3},
  pages = {29--39},
  issn = {00223115},
  doi = {10.1016/j.jnucmat.2013.05.012},
  urldate = {2025-05-19},
  langid = {english}
}

@article{cococcioniLinearResponseApproach2005a,
  title = {Linear Response Approach to the Calculation of the Effective Interaction Parameters in the {{LDA}}+{{U}} Method},
  author = {Cococcioni, Matteo and De Gironcoli, Stefano},
  year = {2005},
  month = jan,
  journal = {Phys. Rev. B},
  volume = {71},
  number = {3},
  pages = {035105},
  issn = {1098-0121, 1550-235X},
  doi = {10.1103/PhysRevB.71.035105},
  urldate = {2025-05-19},
  copyright = {http://link.aps.org/licenses/aps-default-license},
  langid = {english}
}

@article{ewingGeologicalDisposalNuclear2016,
  title = {Geological {{Disposal}} of {{Nuclear Waste}}: A {{Primer}}},
  shorttitle = {Geological {{Disposal}} of {{Nuclear Waste}}},
  author = {Ewing, Rodney C. and Whittleston, Robert A. and Yardley, Bruce W.D.},
  year = {2016},
  month = aug,
  journal = {ELEMENTS},
  volume = {12},
  number = {4},
  pages = {233--237},
  issn = {1811-5209, 1811-5217},
  doi = {10.2113/gselements.12.4.233},
  urldate = {2025-05-19},
  langid = {english}
}

@article{giannozziAdvancedCapabilitiesMaterials2017b,
  title = {Advanced Capabilities for Materials Modelling with {{Quantum ESPRESSO}}},
  author = {Giannozzi, P and Andreussi, O and Brumme, T and Bunau, O and Buongiorno Nardelli, M and Calandra, M and Car, R and Cavazzoni, C and Ceresoli, D and Cococcioni, M and Colonna, N and Carnimeo, I and Dal Corso, A and De Gironcoli, S and Delugas, P and DiStasio, R A and Ferretti, A and Floris, A and Fratesi, G and Fugallo, G and Gebauer, R and Gerstmann, U and Giustino, F and Gorni, T and Jia, J and Kawamura, M and Ko, H-Y and Kokalj, A and K{\"u}{\c c}{\"u}kbenli, E and Lazzeri, M and Marsili, M and Marzari, N and Mauri, F and Nguyen, N L and Nguyen, H-V and {Otero-de-la-Roza}, A and Paulatto, L and Ponc{\'e}, S and Rocca, D and Sabatini, R and Santra, B and Schlipf, M and Seitsonen, A P and Smogunov, A and Timrov, I and Thonhauser, T and Umari, P and Vast, N and Wu, X and Baroni, S},
  year = {2017},
  month = nov,
  journal = {J. Phys.: Condens. Matter},
  volume = {29},
  number = {46},
  pages = {465901},
  issn = {0953-8984, 1361-648X},
  doi = {10.1088/1361-648X/aa8f79},
  urldate = {2025-05-19}
}

@article{Giannozzi:2020,
  author  = {Giannozzi, P. and Baseggio, O. and Bonf\`a, P. and Brunato, D. and Car, R. and Carnimeo, I. and Cavazzoni, C. and de~Gironcoli, S. and Delugas, P. and Ferrari~Ruffino, F. and Ferretti, A. and Marzari, N. and Timrov, I. and Urru, A. and Baroni, S.},
  title   = {{Quantum ESPRESSO toward the exascale}},
  journal = {J. Chem. Phys.},
  volume  = {152},
  pages   = {154105},
  year    = {2020},
  doi     = {10.1063/5.0005082}
}

@article{giannozziQUANTUMESPRESSOModular2009b,
  title = {{{QUANTUM ESPRESSO}}: A Modular and Open-Source Software Project for Quantum Simulations of Materials},
  shorttitle = {{{QUANTUM ESPRESSO}}},
  author = {Giannozzi, Paolo and Baroni, Stefano and Bonini, Nicola and Calandra, Matteo and Car, Roberto and Cavazzoni, Carlo and Ceresoli, Davide and Chiarotti, Guido L and Cococcioni, Matteo and Dabo, Ismaila and Dal Corso, Andrea and De Gironcoli, Stefano and Fabris, Stefano and Fratesi, Guido and Gebauer, Ralph and Gerstmann, Uwe and Gougoussis, Christos and Kokalj, Anton and Lazzeri, Michele and {Martin-Samos}, Layla and Marzari, Nicola and Mauri, Francesco and Mazzarello, Riccardo and Paolini, Stefano and Pasquarello, Alfredo and Paulatto, Lorenzo and Sbraccia, Carlo and Scandolo, Sandro and Sclauzero, Gabriele and Seitsonen, Ari P and Smogunov, Alexander and Umari, Paolo and Wentzcovitch, Renata M},
  year = {2009},
  month = sep,
  journal = {J. Phys.: Condens. Matter},
  volume = {21},
  number = {39},
  pages = {395502},
  issn = {0953-8984, 1361-648X},
  doi = {10.1088/0953-8984/21/39/395502},
  urldate = {2025-05-19}
}

@article{heLowspinStateFe2023,
  title = {Low-Spin State of {{Fe}} in {{Fe-doped NiOOH}} Electrocatalysts},
  author = {He, Zheng-Da and Tesch, Rebekka and Eslamibidgoli, Mohammad J. and Eikerling, Michael H. and Kowalski, Piotr M.},
  year = {2023},
  month = jun,
  journal = {Nat. Commun.},
  volume = {14},
  number = {1},
  pages = {3498},
  issn = {2041-1723},
  doi = {10.1038/s41467-023-38978-5},
  urldate = {2025-05-19},
  langid = {english}
}

@article{kowalskiElectrodeElectrolyteMaterials2021,
  title = {Electrode and {{Electrolyte Materials From Atomistic Simulations}}: {{Properties}} of {{Li{\textsubscript{x}}FePO{\textsubscript{4}} Electrode}} and {{Zircon-Based Ionic Conductors}}},
  shorttitle = {Electrode and {{Electrolyte Materials From Atomistic Simulations}}},
  author = {Kowalski, Piotr M. and He, Zhengda and Cheong, Oskar},
  year = {2021},
  month = mar,
  journal = {Front. Energy Res.},
  volume = {9},
  pages = {653542},
  issn = {2296-598X},
  doi = {10.3389/fenrg.2021.653542},
  urldate = {2025-05-19}
}

@article{kulikSystematicStudyFirstrow2010a,
  title = {Systematic Study of First-Row Transition-Metal Diatomic Molecules: {{A}} Self-Consistent {{DFT}}+{{U}} Approach},
  shorttitle = {Systematic Study of First-Row Transition-Metal Diatomic Molecules},
  author = {Kulik, Heather J. and Marzari, Nicola},
  year = {2010},
  month = sep,
  journal = {J. Chem. Phys.},
  volume = {133},
  number = {11},
  pages = {114103},
  issn = {0021-9606, 1089-7690},
  doi = {10.1063/1.3489110},
  urldate = {2025-05-19},
  langid = {english}
}

@article{kvashninaTrendsValenceBand2018,
  title = {Trends in the Valence Band Electronic Structures of Mixed Uranium Oxides},
  author = {Kvashnina, Kristina O. and Kowalski, Piotr M. and Butorin, Sergei M. and Leinders, Gregory and Pakarinen, Janne and B{\`e}s, Ren{\'e} and Li, Haijian and Verwerft, Marc},
  year = {2018},
  journal = {Chem. Commun.},
  volume = {54},
  number = {70},
  pages = {9757--9760},
  issn = {1359-7345, 1364-548X},
  doi = {10.1039/C8CC05464A},
  urldate = {2025-05-19},
  langid = {english}
}

@article{mackeOrbitalResolvedDFTMolecules2024,
  title = {Orbital-{{Resolved DFT}}{\emph{+}}{{{\emph{U}}}} for {{Molecules}} and {{Solids}}},
  author = {Macke, Eric and Timrov, Iurii and Marzari, Nicola and Ciacchi, Lucio Colombi},
  year = {2024},
  month = jun,
  journal = {J. Chem. Theory Comput.},
  volume = {20},
  number = {11},
  pages = {4824--4843},
  issn = {1549-9618, 1549-9626},
  doi = {10.1021/acs.jctc.3c01403},
  urldate = {2025-05-19},
  copyright = {https://creativecommons.org/licenses/by/4.0/},
  langid = {english}
}

@article{murphyHighPressureSynthesisStructural2018,
  title = {High-{{Pressure Synthesis}}, {{Structural}}, and {{Spectroscopic Studies}} of the {{Ni}}--{{U}}--{{O System}}},
  author = {Murphy, Gabriel L. and Kegler, Philip and Zhang, Yingjie and Zhang, Zhaoming and Alekseev, Evgeny V. and De Jonge, Martin D. and Kennedy, Brendan J.},
  year = {2018},
  month = nov,
  journal = {Inorg. Chem.},
  volume = {57},
  number = {21},
  pages = {13847--13858},
  issn = {0020-1669, 1520-510X},
  doi = {10.1021/acs.inorgchem.8b02355},
  urldate = {2025-05-19},
  langid = {english}
}

@article{murphyTiltingDistortionRutileRelated2021,
  title = {Tilting and {{Distortion}} in {{Rutile-Related Mixed Metal Ternary Uranium Oxides}}: {{A Structural}}, {{Spectroscopic}}, and {{Theoretical Investigation}}},
  shorttitle = {Tilting and {{Distortion}} in {{Rutile-Related Mixed Metal Ternary Uranium Oxides}}},
  author = {Murphy, Gabriel L. and Zhang, Zhaoming and Tesch, Rebekka and Kowalski, Piotr M. and Avdeev, Maxim and Kuo, Eugenia Y. and Gregg, Daniel J. and Kegler, Philip and Alekseev, Evgeny V. and Kennedy, Brendan J.},
  year = {2021},
  month = feb,
  journal = {Inorg. Chem.},
  volume = {60},
  number = {4},
  pages = {2246--2260},
  issn = {0020-1669, 1520-510X},
  doi = {10.1021/acs.inorgchem.0c03077},
  urldate = {2025-05-19},
  copyright = {https://doi.org/10.15223/policy-029},
  langid = {english}
}

@article{tingRefinedDFTMethod2023,
  title = {Refined {{DFT}}+{{U}} Method for Computation of Layered Oxide Cathode Materials},
  author = {Ting, Yin-Ying and Kowalski, Piotr M.},
  year = {2023},
  month = mar,
  journal = {Electrochim. Acta},
  volume = {443},
  pages = {141912},
  issn = {00134686},
  doi = {10.1016/j.electacta.2023.141912},
  urldate = {2025-05-19},
  langid = {english}
}

@article{vitovaDehydrationUranylPeroxide2018,
  title = {Dehydration of the {{Uranyl Peroxide Studtite}}, [{{UO}}{\textsubscript{2}}({$\eta$}{\textsuperscript{2}}-{{O}}{\textsubscript{2}})({{H}}{\textsubscript{2}}{{O}}){\textsubscript{2}}]{$\cdot$}{{2H}}{\textsubscript{2}}{{O}}, {{Affords}} a {{Drastic Change}} in the {{Electronic Structure}}: {{A Combined X-ray Spectroscopic}} and {{Theoretical Analysis}}},
  shorttitle = {Dehydration of the {{Uranyl Peroxide Studtite}}, [{{UO}}{\textsubscript{2}} ({$\eta$}{\textsuperscript{2}} -{{O}}{\textsubscript{2}} )({{H}}{\textsubscript{2}} {{O}}){\textsubscript{2}} ]{$\cdot$}{{2H}}{\textsubscript{2}} {{O}}, {{Affords}} a {{Drastic Change}} in the {{Electronic Structure}}},
  author = {Vitova, Tonya and Pidchenko, Ivan and Biswas, Saptarshi and Beridze, George and Dunne, Peter W. and Schild, Dieter and Wang, Zheming and Kowalski, Piotr M. and Baker, Robert J.},
  year = {2018},
  month = feb,
  journal = {Inorg. Chem.},
  volume = {57},
  number = {4},
  pages = {1735--1743},
  issn = {0020-1669, 1520-510X},
  doi = {10.1021/acs.inorgchem.7b02326},
  urldate = {2025-05-19},
  langid = {english}
}

@article{lowdinNonOrthogonalityProblemConnected1950,
  title = {On the {{Non}}-{{Orthogonality Problem Connected}} with the {{Use}} of {{Atomic Wave Functions}} in the {{Theory}} of {{Molecules}} and {{Crystals}}},
  author = {L{\"o}wdin, Per-Olov},
  year = {1950},
  month = mar,
  journal = {J. Chem. Phys.},
  volume = {18},
  number = {3},
  pages = {365--375},
  issn = {0021-9606, 1089-7690},
  doi = {10.1063/1.1747632},
  urldate = {2022-12-15},
  langid = {english}
}

@article{Timrov:2018,
  author  = {I.~Timrov and N.~Marzari and M.~Cococcioni},
  title   = {Hubbard parameters from density-functional perturbation theory},
  journal = {Phys. Rev. B},
  volume  = {98},
  pages   = {085127},
  year    = {2018},
  DOI = {10.1103/PhysRevB.98.085127}
}

@article{timrovPulayForcesDensityfunctional2020,
  title = {Pulay Forces in Density-Functional Theory with Extended {{Hubbard}} Functionals: {{From}} Nonorthogonalized to Orthogonalized Manifolds},
  shorttitle = {Pulay Forces in Density-Functional Theory with Extended {{Hubbard}} Functionals},
  author = {Timrov, Iurii and Aquilante, Francesco and Binci, Luca and Cococcioni, Matteo and Marzari, Nicola},
  year = {2020},
  month = dec,
  journal = {Phys. Rev. B},
  volume = {102},
  number = {23},
  pages = {235159},
  issn = {2469-9950, 2469-9969},
  doi = {10.1103/PhysRevB.102.235159},
  urldate = {2022-12-05},
  langid = {english}
}

@article{perdewSelfinteractionCorrection1981a,
  title = {Self-Interaction Correction to Density-Functional Approximations for Many-Electron Systems},
  author = {Perdew, J. P. and Zunger, Alex},
  year = {1981},
  month = may,
  journal = {Phys. Rev. B},
  volume = {23},
  number = {10},
  pages = {5048--5079},
  publisher = {American Physical Society},
  doi = {10.1103/PhysRevB.23.5048},
  urldate = {2025-02-10}
}

@article{pickettReformulationLDAMethod1998,
  title = {Reformulation of the {{LDA}} + {{U}} Method for a Local-Orbital Basis},
  author = {Pickett, W. E. and Erwin, S. C. and Ethridge, E. C.},
  year = {1998},
  month = jul,
  journal = {Phys. Rev. B},
  volume = {58},
  number = {3},
  pages = {1201--1209},
  issn = {0163-1829, 1095-3795},
  doi = {10.1103/PhysRevB.58.1201},
  urldate = {2023-01-12},
  langid = {english}
}

@article{solovyev2All31996,
  title = {T\textsubscript{2g} versus All 3{\emph{d}} Localization in {{La}}{{{\emph{M}}}}{{O}}\textsubscript{3} Perovskites ({{{\emph{M}}}}={{Ti}}--{{Cu}}): {{First-principles}} Study},
  shorttitle = {T 2 g versus All 3 {\emph{d}} Localization in {{La}} {{{\emph{M}}}} {{O}} 3 Perovskites ( {{{\emph{M}}}} ={{Ti}}--{{Cu}})},
  author = {Solovyev, Igor and Hamada, Noriaki and Terakura, Kiyoyuki},
  year = {1996},
  month = mar,
  journal = {Phys. Rev. B},
  volume = {53},
  number = {11},
  pages = {7158--7170},
  issn = {0163-1829, 1095-3795},
  doi = {10.1103/PhysRevB.53.7158},
  urldate = {2023-07-13},
  langid = {english}
}

@article{sitSimple2011,
  title = {Simple, {{Unambiguous Theoretical Approach}} to {{Oxidation State Determination}} via {{First-Principles Calculations}}},
  author = {Sit, Patrick H.-L. and Car, Roberto and Cohen, Morrel H. and Selloni, Annabella},
  year = {2011},
  month = oct,
  journal = {Inorg. Chem.},
  volume = {50},
  number = {20},
  pages = {10259--10267},
  issn = {0020-1669, 1520-510X},
  doi = {10.1021/ic2013107},
  urldate = {2022-04-20},
  langid = {english}
}

@article{perdewRestoringDensityGradientExpansion2008a,
  title = {Restoring the {{Density-Gradient Expansion}} for {{Exchange}} in {{Solids}} and {{Surfaces}}},
  author = {Perdew, John P. and Ruzsinszky, Adrienn and Csonka, G{\'a}bor I. and Vydrov, Oleg A. and Scuseria, Gustavo E. and Constantin, Lucian A. and Zhou, Xiaolan and Burke, Kieron},
  year = {2008},
  month = apr,
  journal = {Phys. Rev. Lett.},
  volume = {100},
  number = {13},
  pages = {136406},
  issn = {0031-9007, 1079-7114},
  doi = {10.1103/PhysRevLett.100.136406},
  urldate = {2025-05-20},
  copyright = {http://link.aps.org/licenses/aps-default-license},
  langid = {english}
}

@article{vanderbiltSoftSelfconsistentPseudopotentials1990,
  title = {Soft Self-Consistent Pseudopotentials in a Generalized Eigenvalue Formalism},
  author = {Vanderbilt, David},
  year = {1990},
  month = apr,
  journal = {Phys. Rev. B},
  volume = {41},
  number = {11},
  pages = {7892--7895},
  issn = {0163-1829, 1095-3795},
  doi = {10.1103/PhysRevB.41.7892},
  urldate = {2025-05-20},
  copyright = {http://link.aps.org/licenses/aps-default-license},
  langid = {english}
}

@article{kulikDensityFunctionalTheory2006,
  title = {Density {{Functional Theory}} in {{Transition-Metal Chemistry}}: {{A Self-Consistent Hubbard \textit{U} Approach}}},
  shorttitle = {Density {{Functional Theory}} in {{Transition-Metal Chemistry}}},
  author = {Kulik, Heather J. and Cococcioni, Matteo and Scherlis, Damian A. and Marzari, Nicola},
  year = {2006},
  month = sep,
  journal = {Phys. Rev. Lett.},
  volume = {97},
  number = {10},
  pages = {103001},
  issn = {0031-9007, 1079-7114},
  doi = {10.1103/PhysRevLett.97.103001},
  urldate = {2023-11-07},
  langid = {english}
}

@article{dudarevElectronenergylossSpectraStructural1998,
  title = {Electron-Energy-Loss Spectra and the Structural Stability of Nickel Oxide: {{An LSDA}}+{{U}} Study},
  shorttitle = {Electron-Energy-Loss Spectra and the Structural Stability of Nickel Oxide},
  author = {Dudarev, S. L. and Botton, G. A. and Savrasov, S. Y. and Humphreys, C. J. and Sutton, A. P.},
  year = {1998},
  month = jan,
  journal = {Phys. Rev. B},
  volume = {57},
  number = {3},
  pages = {1505--1509},
  issn = {0163-1829, 1095-3795},
  doi = {10.1103/PhysRevB.57.1505},
  urldate = {2020-05-25},
  langid = {english}
}

@article{uhrinMachineLearningHubbard2025,
  title = {Machine Learning {{Hubbard}} Parameters with Equivariant Neural Networks},
  author = {Uhrin, Martin and Zadoks, Austin and Binci, Luca and Marzari, Nicola and Timrov, Iurii},
  year = {2025},
  month = jan,
  journal = {npj. Comput. Mater.},
  volume = {11},
  number = {1},
  pages = {19},
  issn = {2057-3960},
  doi = {10.1038/s41524-024-01501-5},
  urldate = {2025-05-20},
  langid = {english}
}

@article{bastoneroFirstprinciplesHubbardParameters2025,
  title = {First-Principles {{Hubbard}} Parameters with Automated and Reproducible Workflows},
  author = {Bastonero, Lorenzo and Malica, Cristiano and Macke, Eric and Bercx, Marnik and Huber, Sebastian P. and Timrov, Iurii and Marzari, Nicola},
  journal = {npj Comput. Mater.},
  volume = {11},
  pages = {183},
  year = {2025},
  doi={10.1038/s41524-025-01685-4},
}

@article{vaugierHubbardHundExchange2012,
  title = {Hubbard {{U}} and {{Hund}} Exchange {{J}} in Transition Metal Oxides: {{Screening}} versus Localization Trends from Constrained Random Phase Approximation},
  shorttitle = {Hubbard {{U}} and {{Hund}} Exchange {{J}} in Transition Metal Oxides},
  author = {Vaugier, Lo{\"i}g and Jiang, Hong and Biermann, Silke},
  year = {2012},
  month = oct,
  journal = {Phys. Rev. B},
  volume = {86},
  number = {16},
  pages = {165105},
  issn = {1098-0121, 1550-235X},
  doi = {10.1103/PhysRevB.86.165105},
  urldate = {2023-05-29},
  langid = {english}
}

@article{leiria_campo_jr_extended_2010,
    title = {Extended {DFT}+\textit{{U}}+\textit{{V}} method with on-site and inter-site electronic interactions},
    volume = {22},
    issn = {0953-8984, 1361-648X},
    url = {https://iopscience.iop.org/article/10.1088/0953-8984/22/5/055602},
    doi = {10.1088/0953-8984/22/5/055602},
    number = {5},
    urldate = {2020-05-25},
    journal = {J. Phys.:Condens. Matter},
    author = {Leiria Campo Jr, Vivaldo and Cococcioni, Matteo},
    month = feb,
    year = {2010},
    pages = {055602},
}

@article{mariano_biased_2020,
    title = {Biased {Spin}-{State} {Energetics} of {Fe}({II}) {Molecular} {Complexes} within {Density}-{Functional} {Theory} and the {Linear}-{Response} {Hubbard} \textit{{U}} {Correction}},
    volume = {16},
    issn = {1549-9618, 1549-9626},
    url = {https://pubs.acs.org/doi/10.1021/acs.jctc.0c00628},
    doi = {10.1021/acs.jctc.0c00628},
    language = {en},
    number = {11},
    urldate = {2022-04-20},
    journal = {J. Chem. Theory Comput.},
    author = {Mariano, Lorenzo A. and Vlaisavljevich, Bess and Poloni, Roberta},
    month = nov,
    year = {2020},
    pages = {6755--6762},
}

@article{mahajanImportance2021,
  title = {Importance of Intersite {{Hubbard}} Interactions in {$\beta-$}{{MnO2}}: {{A}} First-Principles {{DFT}}+{{U}}+{{V}} Study},
  shorttitle = {Importance of Intersite {{Hubbard}} Interactions in {$\beta$} - {{MnO}} 2},
  author = {Mahajan, Ruchika and Timrov, Iurii and Marzari, Nicola and Kashyap, Arti},
  year = {2021},
  month = oct,
  journal = {Phys. Rev. Mater.},
  volume = {5},
  number = {10},
  pages = {104402},
  issn = {2475-9953},
  doi = {10.1103/PhysRevMaterials.5.104402},
  urldate = {2023-01-06},
  langid = {english}
}

@article{kummelOrbitaldependentDensityFunctionals2008,
  title = {Orbital-Dependent Density Functionals: {{Theory}} and Applications},
  shorttitle = {Orbital-Dependent Density Functionals},
  author = {K{\"u}mmel, Stephan and Kronik, Leeor},
  year = {2008},
  month = jan,
  journal = {Rev. Mod. Phys.},
  volume = {80},
  number = {1},
  pages = {3--60},
  publisher = {American Physical Society},
  doi = {10.1103/RevModPhys.80.3},
  urldate = {2025-03-12}
}

@article{kronikPiecewiseLinearityFreedom2020,
  title = {Piecewise Linearity, Freedom from Self-Interaction, and a {{Coulomb}} Asymptotic Potential: Three Related yet Inequivalent Properties of the Exact Density Functional},
  shorttitle = {Piecewise Linearity, Freedom from Self-Interaction, and a {{Coulomb}} Asymptotic Potential},
  author = {Kronik, Leeor and K{\"u}mmel, Stephan},
  year = {2020},
  journal = {Phys. Chem. Chem. Phys.},
  volume = {22},
  number = {29},
  pages = {16467--16481},
  issn = {1463-9076, 1463-9084},
  doi = {10.1039/D0CP02564J},
  urldate = {2025-06-12},
  langid = {english}
}

@article{mori-sanchez_many-electron_2006,
    title = {Many-electron self-interaction error in approximate density functionals},
    volume = {125},
    issn = {0021-9606},
    url = {https://doi.org/10.1063/1.2403848},
    doi = {10.1063/1.2403848},
    number = {20},
    urldate = {2025-02-10},
    journal = {J. Chem. Phys.},
    author = {Mori-Sánchez, Paula and Cohen, Aron J. and Yang, Weitao},
    month = nov,
    year = {2006},
    pages = {201102},
}

@article{zhao_global_2016,
    title = {Global and local curvature in density functional theory},
    volume = {145},
    issn = {0021-9606, 1089-7690},
    url = {http://aip.scitation.org/doi/10.1063/1.4959882},
    doi = {10.1063/1.4959882},
    language = {en},
    number = {5},
    urldate = {2022-12-05},
    journal = {J. Chem. Phys.},
    author = {Zhao, Qing and Ioannidis, Efthymios I. and Kulik, Heather J.},
    month = aug,
    year = {2016},
    pages = {054109},
}

@article{anisimov_band_1991,
    title = {Band theory and {Mott} insulators: {Hubbard} \textit{{U}} instead of {Stoner} \textit{{I}}},
    volume = {44},
    issn = {0163-1829, 1095-3795},
    shorttitle = {Band theory and {Mott} insulators},
    url = {https://link.aps.org/doi/10.1103/PhysRevB.44.943},
    doi = {10.1103/PhysRevB.44.943},
    language = {en},
    number = {3},
    urldate = {2023-11-07},
    journal = {Phys. Rev. B},
    author = {Anisimov, Vladimir I. and Zaanen, Jan and Andersen, Ole K.},
    month = jul,
    year = {1991},
    pages = {943--954},
}

@article{solovyev_corrected_1994,
    title = {Corrected atomic limit in the local-density approximation and the electronic structure of \textit{d} impurities in {Rb}},
    volume = {50},
    issn = {0163-1829, 1095-3795},
    url = {https://link.aps.org/doi/10.1103/PhysRevB.50.16861},
    doi = {10.1103/PhysRevB.50.16861},
    language = {en},
    number = {23},
    urldate = {2024-03-05},
    journal = {Phys. Rev. B},
    author = {Solovyev, I. V. and Dederichs, P. H. and Anisimov, V. I.},
    month = dec,
    year = {1994},
    pages = {16861--16871},
}

@article{liechtenstein_density-functional_1995,
    title = {Density-functional theory and strong interactions: {Orbital} ordering in {Mott}-{Hubbard} insulators},
    volume = {52},
    issn = {0163-1829, 1095-3795},
    shorttitle = {Density-functional theory and strong interactions},
    url = {https://link.aps.org/doi/10.1103/PhysRevB.52.R5467},
    doi = {10.1103/PhysRevB.52.R5467},
    language = {en},
    number = {8},
    urldate = {2023-11-07},
    journal = {Phys. Rev. B},
    author = {Liechtenstein, A. I. and Anisimov, V. I. and Zaanen, J.},
    month = aug,
    year = {1995},
    pages = {R5467--R5470},
}

@article{bajaj_communication_2017,
    title = {Communication: {Recovering} the flat-plane condition in electronic structure theory at semi-local {DFT} cost},
    volume = {147},
    issn = {0021-9606, 1089-7690},
    shorttitle = {Communication},
    url = {http://aip.scitation.org/doi/10.1063/1.5008981},
    doi = {10.1063/1.5008981},
    language = {en},
    number = {19},
    urldate = {2022-12-06},
    journal = {J. Chem. Phys.},
    author = {Bajaj, Akash and Janet, Jon Paul and Kulik, Heather J.},
    month = nov,
    year = {2017},
    pages = {191101},
}

@article{burgess_dft_2023,
    title = {{DFT}+{U}-type functional derived to explicitly address the flat plane condition},
    volume = {107},
    issn = {2469-9950, 2469-9969},
    url = {https://link.aps.org/doi/10.1103/PhysRevB.107.L121115},
    doi = {10.1103/PhysRevB.107.L121115},
    language = {en},
    number = {12},
    urldate = {2024-03-05},
    journal = {Phys. Rev. B},
    author = {Burgess, Andrew C. and Linscott, Edward and O'Regan, David D.},
    month = mar,
    year = {2023},
    pages = {L121115},
}

@article{zhouObtainingCorrectOrbital2009,
  title = {Obtaining Correct Orbital Ground States in \textit{f}-Electron Systems Using a Nonspherical Self-Interaction-Corrected LDA\textit{+U} Method},
  author = {Zhou, Fei and Ozoli{\c n}{\v s}, V.},
  year = {2009},
  month = sep,
  journal = {Phys. Rev. B},
  volume = {80},
  number = {12},
  pages = {125127},
  publisher = {American Physical Society},
  doi = {10.1103/PhysRevB.80.125127},
  urldate = {2025-06-13}
}

@article{oregan_subspace_2011,
    title = {Subspace representations in \textit{ab initio} methods for strongly correlated systems},
    volume = {83},
    copyright = {http://link.aps.org/licenses/aps-default-license},
    issn = {1098-0121, 1550-235X},
    url = {https://link.aps.org/doi/10.1103/PhysRevB.83.245124},
    doi = {10.1103/PhysRevB.83.245124},
    language = {en},
    number = {24},
    urldate = {2025-06-13},
    journal = {Phys. Rev. B},
    author = {O’Regan, David D. and Payne, Mike C. and Mostofi, Arash A.},
    month = jun,
    year = {2011},
    pages = {245124},
}

@article{timrovSelfconsistentHubbardParameters2021,
  title = {Self-Consistent {{Hubbard}} Parameters from Density-Functional Perturbation Theory in the Ultrasoft and Projector-Augmented Wave Formulations},
  author = {Timrov, Iurii and Marzari, Nicola and Cococcioni, Matteo},
  year = {2021},
  month = jan,
  journal = {Phys. Rev. B},
  volume = {103},
  number = {4},
  pages = {045141},
  issn = {2469-9950, 2469-9969},
  doi = {10.1103/PhysRevB.103.045141},
  urldate = {2025-06-17},
  langid = {english}
}

@article{linscottRoleSpinCalculation2018,
  title = {Role of Spin in the Calculation of {{Hubbard U}} and {{Hund}}'s {{J}} Parameters from First Principles},
  author = {Linscott, Edward B. and Cole, Daniel J. and Payne, Michael C. and O'Regan, David D.},
  year = {2018},
  month = dec,
  journal = {Phys. Rev. B},
  volume = {98},
  number = {23},
  pages = {235157},
  issn = {2469-9950, 2469-9969},
  doi = {10.1103/PhysRevB.98.235157},
  urldate = {2023-05-30},
  langid = {english}
}

@article{cartaExplicitDemonstrationEquivalence2025,
  title = {Explicit Demonstration of the Equivalence between {{DFT}}+{{U}} and the {{Hartree-Fock}} Limit of {{DFT}}+{{DMFT}}},
  author = {Carta, Alberto and Timrov, Iurii and Mlkvik, Peter and Hampel, Alexander and Ederer, Claude},
  year = {2025},
  month = mar,
  journal = {Phys. Rev. Res.},
  volume = {7},
  number = {1},
  publisher = {American Physical Society (APS)},
  issn = {2643-1564},
  doi = {10.1103/physrevresearch.7.013289},
  urldate = {2025-07-08},
  copyright = {https://creativecommons.org/licenses/by/4.0/},
  langid = {english}
}

@article{orhanFirstprinciplesHubbardHunds2020,
  title = {First-Principles {{Hubbard U}} and {{Hund}}'s {{J}} Corrected Approximate Density Functional Theory Predicts an Accurate Fundamental Gap in Rutile and Anatase {{TiO\textsubscript{2}}}},
  author = {Orhan, Okan K. and O'Regan, David D.},
  year = {2020},
  month = jun,
  journal = {Phys. Rev. B},
  volume = {101},
  number = {24},
  publisher = {American Physical Society (APS)},
  issn = {2469-9950, 2469-9969},
  doi = {10.1103/physrevb.101.245137},
  urldate = {2025-07-10},
  copyright = {https://link.aps.org/licenses/aps-default-license},
  langid = {english}
}

@article{shannonRevisedEffectiveIonic1976,
  title = {Revised Effective Ionic Radii and Systematic Studies of Interatomic Distances in Halides and Chalcogenides},
  author = {Shannon, R. D.},
  year = {1976},
  month = sep,
  journal = {Acta Crystallogr., Sect. A},
  volume = {32},
  number = {5},
  pages = {751--767},
  publisher = {International Union of Crystallography (IUCr)},
  issn = {0567-7394},
  doi = {10.1107/s0567739476001551},
  urldate = {2025-07-14},
  copyright = {http://journals.iucr.org/services/copyrightpolicy.html}
}

@article{doDelocalizedPolaronBursteinMoss2023,
  title = {Delocalized Polaron and {{Burstein-Moss}} Shift Induced by {{Li}} in {$\alpha-$}{{V\textsubscript{2}O\textsubscript{5}}}: {{A DFT}}+{{DMFT}} Study},
  shorttitle = {Delocalized Polaron and {{Burstein-Moss}} Shift Induced by {{Li}} in {$\alpha-$}{{V2O5}}},
  author = {Do, Huu T. and Lee, Alex Taekyung and Park, Hyowon and Ngo, Anh T.},
  year = {2023},
  month = nov,
  journal = {Phys. Rev. B},
  volume = {108},
  number = {20},
  publisher = {American Physical Society (APS)},
  issn = {2469-9950, 2469-9969},
  doi = {10.1103/physrevb.108.205122},
  urldate = {2025-07-15},
  copyright = {https://link.aps.org/licenses/aps-default-license},
  langid = {english}
}

@article{liuAnomalousReversalStability2025,
  title = {Anomalous Reversal of Stability in {{Mo-containing}} Oxides: {{A}} Difficult Case Exhibiting Sensitivity to {{DFT}}+{{U}} and Distortion},
  shorttitle = {Anomalous Reversal of Stability in {{Mo-containing}} Oxides},
  author = {Liu, Tzu-chen and Gaines, Dale and Kim, Hyungjun and {Salgado-Casanova}, Adolfo and Torrisi, Steven B. and Wolverton, Chris},
  year = {2025},
  month = may,
  journal = {Phys. Rev. Mater.},
  volume = {9},
  number = {5},
  publisher = {American Physical Society (APS)},
  issn = {2475-9953},
  doi = {10.1103/physrevmaterials.9.055402},
  urldate = {2025-07-15},
  copyright = {https://link.aps.org/licenses/aps-default-license},
  langid = {english}
}

@article{himmetoglu_first-principles_2011,
    title = {First-principles study of electronic and structural properties of {CuO}},
    volume = {84},
    issn = {1098-0121, 1550-235X},
    url = {https://link.aps.org/doi/10.1103/PhysRevB.84.115108},
    doi = {10.1103/PhysRevB.84.115108},
    language = {en},
    number = {11},
    urldate = {2022-12-05},
    journal = {Phys. Rev. B},
    author = {Himmetoglu, Burak and Wentzcovitch, Renata M. and Cococcioni, Matteo},
    month = sep,
    year = {2011},
    pages = {115108},
}

@article{gebreyesusUnderstandingRoleHubbard2023,
  title = {Understanding the Role of {{Hubbard}} Corrections in the Rhombohedral Phase of {{BaTiO\textsubscript{3}}}},
  author = {Gebreyesus, G. and Bastonero, Lorenzo and Kotiuga, Michele and Marzari, Nicola and Timrov, Iurii},
  year = {2023},
  month = dec,
  journal = {Phys. Rev. B},
  volume = {108},
  number = {23},
  publisher = {American Physical Society (APS)},
  issn = {2469-9950, 2469-9969},
  doi = {10.1103/physrevb.108.235171},
  urldate = {2025-07-22},
  copyright = {https://link.aps.org/licenses/aps-default-license},
  langid = {english}
}

@article{qiaoAutomatedMixingMaximally2023,
  title = {Automated Mixing of Maximally Localized {{Wannier}} Functions into Target Manifolds},
  author = {Qiao, Junfeng and Pizzi, Giovanni and Marzari, Nicola},
  year = {2023},
  month = oct,
  journal = {npj. Comput. Mater.},
  volume = {9},
  number = {1},
  publisher = {{Springer Science and Business Media LLC}},
  issn = {2057-3960},
  doi = {10.1038/s41524-023-01147-9},
  urldate = {2025-07-23},
  copyright = {https://creativecommons.org/licenses/by/4.0},
  langid = {english}
}

@article{qiaoProjectabilityDisentanglementAccurate2023,
  title = {Projectability Disentanglement for Accurate and Automated Electronic-Structure {{Hamiltonians}}},
  author = {Qiao, Junfeng and Pizzi, Giovanni and Marzari, Nicola},
  year = {2023},
  month = nov,
  journal = {npj. Comput. Mater.},
  volume = {9},
  number = {1},
  publisher = {{Springer Science and Business Media LLC}},
  issn = {2057-3960},
  doi = {10.1038/s41524-023-01146-w},
  urldate = {2025-07-23},
  copyright = {https://creativecommons.org/licenses/by/4.0},
  langid = {english}
}

@article{himmetoglu_hubbard-corrected_2014,
    title = {Hubbard-corrected {DFT} energy functionals: {The} {LDA}+{U} description of correlated systems},
    volume = {114},
    issn = {00207608},
    shorttitle = {Hubbard-corrected {DFT} energy functionals},
    url = {https://onlinelibrary.wiley.com/doi/10.1002/qua.24521},
    doi = {10.1002/qua.24521},
    language = {en},
    number = {1},
    urldate = {2022-12-05},
    journal = {Int. J. Quantum Chem.},
    author = {Himmetoglu, Burak and Floris, Andrea and de Gironcoli, Stefano and Cococcioni, Matteo},
    month = jan,
    year = {2014},
    pages = {14--49},
}

@article{florisVibrationalPropertiesMnO2011,
  title = {Vibrational Properties of {{MnO}} and {{NiO}} from {{DFT}}+{{U-based}} Density Functional Perturbation Theory},
  author = {Floris, A. and De Gironcoli, S. and Gross, E. K. U. and Cococcioni, M.},
  year = {2011},
  month = oct,
  journal = {Phys. Rev. B},
  volume = {84},
  number = {16},
  publisher = {American Physical Society (APS)},
  issn = {1098-0121, 1550-235X},
  doi = {10.1103/physrevb.84.161102},
  urldate = {2025-07-24},
  copyright = {http://link.aps.org/licenses/aps-default-license},
  langid = {english}
}

@article{cococcioniEnergetics2019,
  title = {Energetics and Cathode Voltages of {{LiMPO\textsubscript{4}}} Olivines ({{M}}={{Fe}},{{Mn}}) from Extended {{Hubbard}} Functionals},
  author = {Cococcioni, Matteo and Marzari, Nicola},
  year = {2019},
  month = mar,
  journal = {Phys. Rev. Mater.},
  volume = {3},
  number = {3},
  pages = {033801},
  issn = {2475-9953},
  doi = {10.1103/PhysRevMaterials.3.033801},
  urldate = {2022-12-05},
  langid = {english}
}

\end{document}